\documentclass[12pt,titlepage]{article}
\usepackage{amsmath}
\usepackage{amssymb}
\usepackage{graphicx}
\usepackage{caption2}
\usepackage{cite}

\newcommand{\be}{\begin{equation}}
\newcommand{\ee}{\end{equation}}
\newcommand{\bea}{\begin{eqnarray}}
\newcommand{\eea}{\end{eqnarray}}
\newcommand{\bef}{\begin{figure}}
\newcommand{\ef}{\end{figure}}
\newcommand{\bt}{\begin{tabular}}
\newcommand{\et}{\end{tabular}}
\newcommand{\bno}{\begin{enumerate}}
\newcommand{\eno}{\end{enumerate}}

\def\3{\ss}

\catcode`\"=\active
\def"{\accent'177}

\begin{document}

\vspace{1.0cm}

 \hspace{3.75cm} {\em ``If the proof is correct, then no other recognition is needed'' 
 
 \hspace{10.0cm} by Grigori Y. Perelman}
 
 \vspace{0.25cm}

\begin{center}
{\bf \Large  On the homotopy multiple-variable method \\and its applications in  the interactions \\ of  nonlinear gravity waves}

\vspace{0.5cm}

Shi-Jun LIAO

State Key Laboratory of Ocean Engineering\\
School of Naval Architecture, Ocean and  Civil Engineering\\
Shanghai Jiao Tong University, Shanghai 200240, China

 (  Email address: sjliao@sjtu.edu.cn  )

\end{center}

\setlength{\parindent}{0.0cm}

{\bf Abstract}  {\em The basic ideas of a homotopy-based multiple-variable method is proposed and applied to investigate the nonlinear interactions of periodic traveling  waves.   Mathematically, this method  does not depend upon any small physical parameters at all and thus is more general than the traditional multiple-scale perturbation techniques.   Physically,  it is found  that, for a fully developed wave system,  the amplitudes of all wave components are finite even if the wave resonance condition given by Phillips (1960)  is exactly satisfied.   Besides, it is revealed that there exist multiple  resonant waves, and that the amplitudes of resonant wave may be much smaller than those of primary waves so that the resonant waves sometimes contain rather small part of wave energy.    Furthermore, a wave resonance condition for arbitrary numbers of  traveling waves with large wave amplitudes is given, which logically contains Phillips' four-wave resonance condition but opens a way to investigate the strongly nonlinear interaction of more than four traveling  waves with large amplitudes.  This work also illustrates that the homotopy multiple-variable method is helpful to gain solutions with important physical meanings of nonlinear problems, if the multiple variables are properly defined with clear physical meanings. }  


\setlength{\parindent}{0.75cm}

\section{Introduction}

In his pioneering work about nonlinear interaction of four gravity waves in deep water, Phillips \cite{Phillips1960JFM} gave the criterion condition for wave resonance
\begin{equation}
{\bf k}_1 \pm {\bf k}_2 \pm {\bf k}_3 \pm {\bf k}_4=0,\;\;\;  \sigma_1 \pm \sigma_2 \pm \sigma_3 \pm\sigma_4 =0, \label{resonance:Phillips}
\end{equation} 
where  $\sigma_i = \sqrt{g k_i}$ with $k_i = |{\bf k}_i|$ ($i=1,2,3,4$) is the wave angular frequency, and ${\bf k}_i$ denotes the wave number.   Note that,  Phillips resonance condition (\ref{resonance:Phillips}) works only for weakly nonlinear waves with small amplitudes,    because  $\sigma_i = \sqrt{g k_i}$  is the angular frequency of linear theory for a single gravity wave with small amplitude.    

Especially, in case of ${\bf k}_4={\bf k}_1$,   Phillips \cite{Phillips1960JFM} showed that, if ${\bf k}_3 = 2{\bf k }_1-{\bf k_2}$ and $\sigma_3 = 2\sigma_1-\sigma_2$, then  a steady-state solution for the triad did not exist,  and the amplitude of the third wave, if initially zero, would grow linearly in time.  This conclusion was confirmed  by Longuet-Higgins \cite{LonguetHiggins1962JFM}  via perturbation theory and  supported by some experiments \cite{LonguetHiggins1966JFM,McGoldrick1966JFM}.    Besides, Benney \cite{Benney1962JFM} solved the equations governing the time dependence of the resonant modes and studied the energy-sharing mechanism  involved.  

Although half century passed since Phillips'  pioneering work \cite{Phillips1960JFM},  there exist still some open questions about nonlinear interaction between gravity waves.   First,  Bretherton \cite{Bretherton1964JFM}   pointed out that the perturbation scheme used by Phillips breaks down for large time, and suggested by investigating a one-dimensional dispersive wave model that the amplitude of each wave component  should be bounded.     Are the amplitudes of each component waves bounded when the resonance condition is exactly satisfied?    Besides,   based on perturbation method, Phillips \cite{Phillips1960JFM} gave the condition (\ref{resonance:Phillips}) of wave resonance for only {\em four} waves.  What is the resonance condition for more waves with large amplitude?   It seems difficult to apply the perturbation scheme used by Phillips \cite{Phillips1960JFM} and Longuet-Higgins \cite{LonguetHiggins1962JFM} to answer these questions, because the related algebra was daunting and ``extraordinarily tedious'', as mentioned by Phillips \cite{Phillips1981JFM}.   

Perturbation techniques  are powerful analytic tools for nonlinear equations, especially when there indeed exists  a small physical parameter, i.e.  perturbation quantity:  the solution is often expressed in a series of the small physical parameter, and the original nonlinear equation is transformed into a sequence of (mostly linear) sub-problems.    Frankly speaking,  perturbation techniques greatly  enrich our knowledge and deepen our understandings about many nonlinear problems.   It has a golden times before the times of computer when people calculated  mostly by hand and wrote the ``analytic''  results on one page (or a few pages) of paper.   However,  some restrictions of perturbation techniques are well-known.  First of all, it depends too strongly on the small physical parameters, but unfortunately many nonlinear problems have no such kind of small parameters.   Besides, perturbation approximations often break down when the so-called perturbation quantity increases.   All of these disadvantages greatly restrict the applications of perturbation techniques. 

In history, the calculating tools have a deep impression on the mathematical methods.  One example is the development of numerical methods such as the finite element method, the finite difference method, and so on.   Fortunately, we are now in the times of computer: a computer can do more than $10^{17}$ fundamental operations and save a great lot of data in diskette within a seconds!  Besides, there are some powerful symbolic computation software  such as Maple, Mathematica, MathLab and so on.   Today, using these software on a laptop,  one can deduce lengthy  formulas, calculate a rather complicated expression, and save analytic results on diskette in a seconds which might be hundreds of papers long if printed out.  So, when the pencil and paper are  replaced by the keyboard, diskette and CPU of a personal computer,  the {\em revolution}  of  analytic approximation methods  comes,  although it is much later than the revolution  of numerical approximation methods.           

What kind of analytic approximation methods should we have in the times of computer?  Can we overcome the restrictions of the traditional perturbation methods?  The answer is rather positive and stimulant.    Note that, the traditional concept of ``analytic'' solution came into being in the times of pen and paper, and it was traditionally believed that a ``analytic'' result must be short and should be written on less than one paper.   Today, it needs only a few seconds to save a rather complicated formula in diskette which might be even more than hundreds of papers.   Besides, one needs only a few seconds to calculate such a lengthy expression by a personal computer.   In fact, one spends much less time to save and calculate such lengthy formula by computer than to calculate and write a simple formula with half-page length by hand.   Therefore, in the times of computer, the concept of ``analytic'' result must be modified, and an ``analytic'' formula can be very lengthy.     Secondly,  due to the high performance of symbolic computation of a computer,  it should be easy to obtain high-order approximations by means of the new analytic methods.      And more importantly, some efficient approaches should be provided to ensure the high accuracy of analytic approximations even for strongly nonlinear problems.  In a short, the new analytic method should belong to the times of computer, and besides should overcome the restrictions of traditional methods mentioned above.   

Such a kind of analytic method has been developed in 1990s when the early symbolic computation software just appeared:  based on the homotopy, a fundamental concept in topology, the homotopy analysis method (HAM) was first proposed by Liao \cite{LiaoBook2003, liao99-nlm, liao99-jfm, liao02-jfm, liao03-jfm, Liao2006-SAM, Liao2007-SAM, Xu2010-PhysicFluids, Li2010-JMP}.  Different from perturbation methods, the HAM has nothing to do with any physical parameters.  More importantly, different from other previous analytic methods,  the HAM provides us a simple way to ensure the convergence of solution series.   Besides, it has been proved that the HAM logically contains the previous non-perturbation methods such as Lyapunov artificial small parameter method, Adomian decomposition method, the $\delta$-expansion method and so on.   Therefore, the HAM is valid for strongly nonlinear problems and  thus is more general.  The HAM has been successfully applied to solve different types of nonlinear ODEs and PDEs in science and finance.

The idea of  multiple-scales of perturbation methods has clear physical meanings.  In this article, based on  the homotopy analysis method (HAM), we propose a multiple-variable technique for general nonlinear differential equations, which keeps the clear physical meaning of multiple-scales but abandons completely the  small physical parameters of perturbation techniques.   In \S 2,  the nonlinear interaction of primary periodic traveling waves in deep water is used as an example to describe the basic ideas of this approach.    In \S 3, we give convergent series solution of a fully developed system of two primary waves even when Phillips' resonance condition is exactly satisfied.   Besides, some interesting results related to multiple resonant waves are reported.  Especially, we reveal that Phillips resonance wave condition is  mathematically equivalent to the zero eigenvalue of the eigenfunction related to the resonant wave. 
 In \S 4.1,  the expression of the eigenvalue of a fully developed wave system for arbitrary number of traveling waves is derived, and then a resonance condition for arbitrary number of small-amplitude waves is given.   In \S 4.2,   a more general resonance condition is further obtained from the physical points of view, which logically contains Phillips' resonance condition (\ref{resonance:Phillips}) but works for arbitrary number  of  traveling waves with large amplitudes.      In \S 5,  the concluding remarks,  open questions and discussions  are given.     The detailed mathematical derivation is given in Appendix A,  and the proof of a convergence theorem is given briefly in Appendix B.

\section{Mathematical description}

Let $z$ denote the vertical co-ordinate, $x,y$ the horizontal co-ordinates,  $t$ the time,  $z=\zeta(x,y,t)$ the free surface,  respectively.   The three axises of $x,y,z$ are perpendicular to each other,  with the unit vector ${\bf i},{\bf j}, {\bf k}$, respectively, i.e.  ${\bf i} \cdot {\bf j} = {\bf i} \cdot {\bf k} = {\bf j} \cdot {\bf k} =0$, where $\cdot$ is the multiplication dot.      Assume that the vorticity is negligible and there exists a potential $\varphi(x,y,z,t)$ for the velocity $\bf u$ that  ${\bf u} = \nabla \varphi$ and
\begin{equation}
\nabla ^2 \varphi = 0,  \;\;\; z\leq \zeta(x,y,t)   \label{geq:Laplace}
\end{equation}
in an incompressible flow, where
\[     \nabla  = {\bf i} \frac{\partial }{\partial x}   +  {\bf j} \frac{\partial }{\partial y} + {\bf k} \frac{\partial }{\partial z}.    \]
 On the free surface $z=\zeta(x,y,t)$, the pressure is constant, which gives from Bernoulli's equation the dynamic boundary condition
 \begin{equation}
 g \; \zeta + \frac{\partial \varphi}{\partial t} +\frac{1}{2} {\bf u}^2 =0,  \;\;\; \mbox{on $z=\zeta(x,y,t)$},\label{bc:zeta}
 \end{equation}
where $g$ is the acceleration due to gravity.   Besides, $z-\zeta$ vanishes following a particle, which gives the kinematic boundary condition
\begin{equation}
 \frac{\partial \zeta}{\partial t} -\frac{\partial \varphi}{\partial z}
 +\left(  \frac{\partial \varphi}{\partial x} \frac{\partial \zeta}{\partial x} + \frac{\partial \varphi}{\partial y} \frac{\partial \zeta}{\partial y}  \right) = 0,\;\;\; \mbox{on $z=\zeta(x,y,t)$}.
 \end{equation} 
 Combining the above two equations gives the boundary condition
 \begin{equation}
 \frac{\partial^2 \varphi}{\partial t^2} +g \frac{\partial \varphi}{\partial z} +\frac{\partial ({\bf u}^2)}{\partial t} + {\bf u}   \cdot  \nabla\left(\frac{1}{2} {\bf u}^2\right) = 0, \;\;\; \mbox{on $z=\zeta(x,y,t)$},  \label{bc:varphi}
 \end{equation}
where  ${\bf u} = \nabla \varphi $ and ${\bf u}^2 = \nabla \varphi  \cdot \nabla \varphi$.   On the bottom, it holds
\begin{equation}
\frac{\partial \varphi}{\partial z} = 0, \;\;\; \mbox{as $z \rightarrow -\infty$}.  \label{bc:varphi:bottom}
\end{equation} 
 For details, please refer to Phillips \cite{Phillips1960JFM}  and Longuet-Higgins \cite{LonguetHiggins1962JFM}  .   
 
Without loss of generality, let us first consider  a  wave system basically composed of  two trains of primary traveling gravity waves in deep water with wave numbers ${\bf k}_1, {\bf k}_2$ and  the corresponding  angular frequencies $\sigma_1,\sigma_2$,   respectively, where ${\bf k}_1 \times  {\bf k}_2 \neq 0 $   (i.e. the two traveling waves are not collinear).  Due to nonlinear interaction, this wave system contains an infinite number of wave components with the corresponding  wave number $m {\bf k}_1 + n{\bf k}_2$, where $m,n$ are integers.    Assume that the wave system is in equilibrium so that each wave amplitude is constant, i.e. independent of the time.  
Let $\alpha_1, \alpha_2$ denote the angles between the positive $x$-axis ${\bf i}$ and the wave number vectors ${\bf k}_1$ and  ${\bf k}_2$, respectively, where ${\bf k}_1 \cdot {\bf k} =  {\bf k}_2 \cdot {\bf k} = 0$, i.e.  the $z$-axis is perpendicular  to the wave numbers  ${\bf k}_1, {\bf k}_2$.   Then, 
 \begin{eqnarray}  
    {\bf k}_1 &=&  k_1 \; \left( \cos \alpha_1  \; {\bf i}  +  \sin\alpha_1 \; {\bf j} \right), \\
    {\bf k}_2 &= & k_2 \; \left( \cos \alpha_2  \; {\bf i}  +   \sin\alpha_2 \; {\bf j} \right), 
    \end{eqnarray}
   where  $ k_1 = |{\bf k}_1|$ and $k_2 = |{\bf k}_2|$.

  Write $ {\bf r} =  x {\bf i} + y {\bf j} $.  According to linear gravity wave theory, the two trains of primary waves traveling  with the wave numbers ${\bf k}_1$ and ${\bf k}_2$ are given by  
 $   a_1 \cos \xi_1,\;  a_2 \cos\xi_2,$
 respectively, where  
 \begin{eqnarray}
     \xi_1 &=&  {\bf k}_1 \cdot {\bf r} - \sigma_1 \; t   = k_1 \cos \alpha_1  \; x +  k_1  \sin\alpha_1 \; y - \sigma_1 \; t\\
     \xi_2 &=& {\bf k}_2 \cdot {\bf r}  - \sigma_2 \; t   = k_2 \cos \alpha_2  \; x +  k_2  \sin\alpha_2 \; y - \sigma_2 \; t.   
  \end{eqnarray}   
So,  the above two variables have very clear physical meaning:  for the considered problem, the wave profile must be a periodic function of $\xi_1$ and $\xi_2$.   Mathematically, using these two variables, the time $t$ should not appear {\em explicitly} for a fully develop wave system.  In other words, one can express the potential function $\varphi(x,y,z,t) = \phi(\xi_1,\xi_2,z)$  and the wave surface $\zeta(x,y,t) = \eta(\xi_1,\xi_2)$  for two trains of primary traveling waves.   In this way, the mathematical expressions of the unknown potential function and wave surface  are clear, with clear physical meanings.

Obviously,   it holds
 \[    \frac{\partial \varphi}{\partial x} = \frac{\partial \phi}{\partial \xi_1} \frac{\partial \xi_1}{\partial x}+  \frac{\partial \phi}{\partial \xi_2}\frac{\partial \xi_2}{\partial x} = \left(  k_1 \cos\alpha_1  \right)  \frac{\partial \phi}{\partial \xi_1}  +  \left( k_2 \cos\alpha_2 \right)  \frac{\partial \phi}{\partial \xi_2} ,\]
 and similarly
  \begin{eqnarray}   
  \frac{\partial \varphi}{\partial y} &=& \left(  k_1 \sin\alpha_1  \right)  \frac{\partial \phi}{\partial \xi_1}  +  \left( k_2 \sin\alpha_2 \right)  \frac{\partial \phi}{\partial \xi_2} , \nonumber\\
  \frac{\partial \varphi}{\partial z} &=&  \frac{\partial \phi}{\partial z}, \nonumber\\
  \frac{\partial \varphi}{\partial t} &=& -\sigma_1 \frac{\partial \phi}{\partial \xi_1} -\sigma_2 \frac{\partial \phi}{\partial \xi_2},\nonumber\\
    \frac{\partial^2 \varphi}{\partial t^2} &=& \sigma_1^2 \; \frac{\partial^2 \phi}{\partial \xi_1^2} + 2 \sigma_1\sigma_2  \; \frac{\partial^2 \phi}{\partial \xi_1 \partial \xi_2} +\sigma_2^2 \; \frac{\partial^2 \phi}{\partial \xi_2^2}. \nonumber 
  \end{eqnarray}
  Then,  
  \begin{eqnarray}
\nabla \varphi &=& {\bf i} \frac{\partial \varphi}{\partial x}+{\bf j} \frac{\partial \varphi}{\partial y} +{\bf k} \frac{\partial \varphi}{\partial z} \nonumber\\
  &=&   k_1 \; \left(\cos \alpha_1  \; {\bf i}  +  \sin\alpha_1 \; {\bf j}\right)  \frac{\partial \phi}{\partial \xi_1} + k_2 \;\left(  \cos \alpha_2  \; {\bf i}  + \sin\alpha_2 \; {\bf j}\right)  \frac{\partial \phi}{\partial \xi_2} +{\bf k} \frac{\partial \phi}{\partial z} \nonumber\\
  &=& {\bf k}_1  \frac{\partial \phi}{\partial \xi_1} +{\bf k}_2  \frac{\partial \phi}{\partial \xi_2}+{\bf k} \frac{\partial \phi}{\partial z} = \hat{\nabla} \phi = {\bf u},
  \end{eqnarray} 
  where
  \begin{equation}
   \hat{\nabla} =  {\bf k}_1  \frac{\partial }{\partial \xi_1} +{\bf k}_2  \frac{\partial}{\partial \xi_2}+{\bf k} \frac{\partial }{\partial z}. \label{def:nabla:hat}
  \end{equation}
  Thus, 
  \begin{eqnarray}
 {\bf u}^2  & = &   \nabla \varphi  \cdot  \nabla \varphi  =  \hat{\nabla} \phi  \cdot \hat{\nabla} \phi \nonumber\\
 &=& k_1^2 \left( \frac{\partial \phi}{\partial \xi_1}\right)^2   + 2 {\bf k}_1 \cdot {\bf k}_2  \; \frac{\partial \phi}{\partial \xi_1}\frac{\partial \phi}{\partial \xi_2}+  k_2^2 \left( \frac{\partial \phi}{\partial \xi_2}\right)^2+\left( \frac{\partial \phi}{\partial z}\right)^2,
\end{eqnarray}
  where ${\bf k}_1 \cdot {\bf k} =  {\bf k}_2 \cdot {\bf k} = 0$ is used, and \[ {\bf k}_1 \cdot {\bf k}_2 =k_1 k_2 \cos(\alpha_1 - \alpha_2). \]
  In general, it holds
  \begin{eqnarray}   
  \hat{\nabla} \phi \cdot  \hat{\nabla} \psi  &= &  k_1^2  \; \frac{\partial \phi}{\partial \xi_1}  \frac{\partial \psi}{\partial \xi_1} + {\bf k}_1 \cdot {\bf k}_2  \; \left( \frac{\partial \phi}{\partial \xi_1}  \frac{\partial \psi}{\partial \xi_2}+ \frac{\partial \psi}{\partial \xi_1}  \frac{\partial \phi}{\partial \xi_2} \right) +  k_2^2  \; \frac{\partial \phi}{\partial \xi_2}  \frac{\partial \psi}{\partial \xi_2}  +  \frac{\partial \phi}{\partial z}  \frac{\partial \psi}{\partial z}\;\;\;\; \;   \label{def:cdot:times}  
 \end{eqnarray}
 for  arbitrary functions $\phi(\xi_1,\xi_2,z)$ and $\psi(\xi_1,\xi_2,z)$.  

 Similarly, we have
 \begin{eqnarray}
     \nabla^2 \varphi &=&  \hat{\nabla}^2 \phi  =  k_1^2  \; \frac{\partial^2 \phi}{\partial \xi_1^2}  + 2 {\bf k}_1 \cdot {\bf k}_2  \; \frac{\partial^2 \phi}{\partial \xi_1\partial \xi_2} + k_2^2 \; \frac{\partial^2 \phi}{\partial \xi_2^2} +\frac{\partial^2 \phi}{\partial z^2}.
 \end{eqnarray}
  Then, the governing equation reads 
 \begin{eqnarray} 
 \hat{\nabla}^2 \phi  =   k_1^2  \; \frac{\partial^2 \phi}{\partial \xi_1^2}  + 2 {\bf k}_1 \cdot {\bf k}_2 \; \frac{\partial^2 \phi}{\partial \xi_1\partial \xi_2} + k_2^2 \; \frac{\partial^2 \phi}{\partial \xi_2^2} +\frac{\partial^2 \phi}{\partial z^2}=0,  \;\;\;\; -\infty < z \leq \eta(\xi_1,\xi_2), \label{geq:phi}
  \end{eqnarray}
  which has the general solution 
  \begin{equation}
  \left[ A \; \cos(m\xi_1 + n \xi_2) +B \; \sin(m\xi_1 + n \xi_2) \right] e^{  |m{\bf k}_1 +  n{\bf k}_2| z },  \label{phi:general}
  \end{equation}
  where $m,n$ are integers and $A, B$ are integral constants.  
  
 For the sake of simplicity, define
 \begin{eqnarray}
    f&=&\frac{1}{2} \hat{\nabla} \phi \cdot  \hat{\nabla}  \phi  = \frac{{\bf u}^2}{2} \nonumber\\
    & =&\frac{1}{2} \left[  k_1^2 \left( \frac{\partial \phi}{\partial \xi_1}\right)^2   + 2 {\bf k}_1 \cdot {\bf k}_2  \; \frac{\partial \phi}{\partial \xi_1}\frac{\partial \phi}{\partial \xi_2}+  k_2^2 \left( \frac{\partial \phi}{\partial \xi_2}\right)^2+\left( \frac{\partial \phi}{\partial z}\right)^2 \right].  
 \end{eqnarray}
  Using the new variables $\xi_1$ and $\xi_2$,   the dynamic boundary condition (\ref{bc:zeta}) becomes
  \begin{equation}
  \eta = \frac{1}{g} \left( \sigma_1\; \frac{\partial \phi}{\partial \xi_1} + \sigma_2 \; \frac{\partial \phi}{\partial \xi_2}  - f \right), \;\;\; \mbox{on $z=\eta(\xi_1,\xi_2)$}.
  \label{bc:eta}
  \end{equation}
  On the free surface $z=\eta(\xi_1,\xi_2)$,  the kinematic boundary condition (\ref{bc:varphi}) reads 
  \begin{eqnarray}
  \sigma_1^2 \; \frac{\partial^2 \phi}{\partial \xi_1^2}+2\sigma_1\sigma_2 \; \frac{\partial^2 \phi}{\partial \xi_1 \partial \xi_2}+\sigma_2^2 \; \frac{\partial^2 \phi}{\partial \xi_2^2}+g \frac{\partial \phi}{\partial z} 
  -2\left( \sigma_1 \;  \frac{\partial f}{\partial \xi_1} + \sigma_2 \; \frac{\partial f}{\partial \xi_2}\right) + \hat{\nabla} \phi \cdot  \hat{\nabla}  f = 0 ,\label{bc:phi}
  \end{eqnarray}
where
\begin{equation}
\frac{\partial f}{\partial \xi_1} = \hat{\nabla} \phi \cdot  \hat{\nabla}  \left(  \frac{\partial \phi}{\partial \xi_1} \right), \;\;  \frac{\partial f}{\partial \xi_2} = \hat{\nabla} \phi \cdot  \hat{\nabla}  \left(  \frac{\partial \phi}{\partial \xi_2} \right) \nonumber
\end{equation}
and
  \begin{eqnarray}   
  \hat{\nabla} \phi \cdot  \hat{\nabla} f  &= &  k_1^2  \; \frac{\partial \phi}{\partial \xi_1}  \frac{\partial f}{\partial \xi_1} + {\bf k}_1 \cdot {\bf k}_2  \; \left( \frac{\partial \phi}{\partial \xi_1}  \frac{\partial f}{\partial \xi_2}+ \frac{\partial f}{\partial \xi_1}  \frac{\partial \phi}{\partial \xi_2} \right) +  k_2^2  \; \frac{\partial \phi}{\partial \xi_2}  \frac{\partial f}{\partial \xi_2}  +  \frac{\partial \phi}{\partial z}  \frac{\partial f}{\partial z}. \nonumber
 \end{eqnarray}
 On the bottom, it holds 
\begin{equation}
\frac{\partial \phi}{\partial z} = 0, \;\;\; \mbox{as $z \rightarrow -\infty$}.  \label{bc:phi:bottom}
\end{equation} 

Given  two angular  frequencies $\sigma_1$ and $\sigma_2$, our aim is to find out  the corresponding unknown potential function $\phi(\xi_1,\xi_2,z)$ and the unknown free surface $\eta(\xi_1,\xi_2)$,  which are governed by  the linear partial differential equation (\ref{geq:phi}) subject to two nonlinear boundary conditions  (\ref{bc:eta}) and (\ref{bc:phi})  on the unknown free surface $z=\eta(\xi_1,\xi_2)$,  and one linear boundary condition (\ref{bc:phi:bottom}) on the bottom.    Here, it should be emphasized that, by means of the two  independent variables $\xi_1$ and $\xi_2$,   the time $t$ does not appear {\em explicitly} in the unknown potential function and wave surface.   More importantly, these new variables have clear physical meanings.   This greatly simplifies  the problem solving, as shown later in this article.

\section{Homotopy-based approach}

As mentioned before,  the two variables $\xi_1$ and $\xi_2$ have clear physical meanings and the solutions of considered problem should be periodic functions of $\xi_1$ and $\xi_2$.    From physical points of view,  it is clear that the wave surface should be in the form
\begin{equation}
\eta(\xi_1,\xi_2) = \sum_{m=0}^{+\infty} \sum_{n= -\infty}^{+\infty} a_{m,n} \cos(m \xi_1+n\xi_2),  \label{def:Solution:Expression:eta}
\end{equation}
where $a_{m,n}$ is the amplitude of wave component $\cos(m\xi_1+n\xi_2)$.    Note that (\ref{phi:general}) is the general solution of the governing equation (\ref{geq:phi}).    So,  the corresponding  potential function should be in the form
\begin{equation}
\phi(\xi_1,\xi_2, z) = \sum_{m=0}^{+\infty} \sum_{n=-\infty}^{+\infty} b_{m,n}  \; \Psi_{m,n}(\xi_1,\xi_2, z)    \label{def:SolutionExpression:phi}
\end{equation}
where
\begin{equation}   \Psi_{m,n}(\xi_1,\xi_2,z)=\sin( m \xi_1 + n \xi_2) \; e^{|m {\bf k}_1 + n {\bf k}_2| z},  \label{def:psi}
\end{equation}
and $b_{m,n}$ is unknown coefficient independent of $\xi_1,\xi_2, z$.     Note that  the potential function  $\phi(\xi_1,\xi_2,z)$ defined by (\ref{def:SolutionExpression:phi})  {\em automatically} satisfies  the governing equation (\ref{geq:phi}).     The above expressions are called the {\em  solution expressions} of $\eta$ and $\phi$,  respectively, which have important role in the frame of the homotopy analysis method.

For simplicity, define a nonlinear operator
  \begin{eqnarray}
{\cal N}\left[  \phi(\xi_1,\xi_2,z)\right] &=& \sigma_1^2 \; \frac{\partial^2 \phi}{\partial \xi_1^2}+2\sigma_1\sigma_2 \; \frac{\partial^2 \phi}{\partial \xi_1 \partial \xi_2}+\sigma_2^2 \; \frac{\partial^2 \phi}{\partial \xi_2^2}+g \frac{\partial \phi}{\partial z} \nonumber\\
 && -2\left( \sigma_1 \;  \frac{\partial f}{\partial \xi_1} + \sigma_2 \; \frac{\partial f}{\partial \xi_2}\right) + \hat{\nabla} \phi \cdot  \hat{\nabla}  f , \label{def:N}
  \end{eqnarray}
where the angular frequencies $\sigma_1,\sigma_2$ are given.   This nonlinear operator is based on the nonlinear boundary condition (\ref{bc:phi}).      Note that it contains a linear operator
\begin{equation}
 {\cal L}_0 \left( \phi \right) =\sigma_1^2 \; \frac{\partial^2 \phi}{\partial \xi_1^2}+2\sigma_1\sigma_2 \; \frac{\partial^2 \phi}{\partial \xi_1 \partial \xi_2}+\sigma_2^2 \; \frac{\partial^2 \phi}{\partial \xi_2^2}+g \frac{\partial \phi}{\partial z}.   \label{def:L[0]}
 \end{equation}
Let  $\cal L$ denote an auxiliary linear differential operator with the property ${\cal L}[0] =0$.  As mentioned in many articles, one of advantages of the homotopy analysis method is the freedom on the choice of the auxiliary linear operator.    Based on the results of linear wave theory, i.e.
\begin{equation}    
 {\sigma}_1 \approx  \sqrt{g\; k_1} = \bar{\sigma}_1, \;\;\;  {\sigma}_2 \approx  \sqrt{ g \; k_2} =\bar{\sigma}_2,      \label{def:sigma:bar}
 \end{equation}
 we {\em choose} such an auxiliary linear operator
\begin{eqnarray}
{\cal L}\phi &=& \bar{\sigma}_{1}^2  \; \frac{\partial^2 \phi}{\partial \xi_1^2} 
+2  \bar{\sigma}_1\bar{\sigma}_2 \; \frac{\partial^2 \phi}{\partial \xi_1 \partial \xi_2}
+\bar{\sigma}_{2}^2 \; \frac{\partial^2 \phi}{\partial \xi_2^2} +g\frac{\partial \phi}{\partial z} .\label{def:L}  
\end{eqnarray}

Then, let $q\in[0,1]$ denote the embedding parameter, $c_0 \neq 0 $  an auxiliary parameter (called convergence-control parameter), $\phi_0(\xi_1,\xi_2, z)$ an initial approximation of the potential function with  $\hat{\nabla} \phi_0 =0$,   respectively.   We construct such a parameterized family of equations (called the zeroth-order deformation equation) in the embedding parameter $q\in[0,1]$:
\begin{equation}
\hat{\nabla}^2 \; \check{\phi}(\xi_1,\xi_2,z;q) = 0,   \;\;\;  -\infty < z \leq \check{\eta}(\xi_1,\xi_2;q) , \label{geq:zero}
\end{equation}
subject to the two boundary conditions on $z = \check{\eta}(\xi_1,\xi_2;q)$:
\begin{equation}
(1-q) \; {\cal L} \left[   \check{\phi}(\xi_1,\xi_2,z;q) -\phi_0(\xi_1,\xi_2,z)   \right] = q \; c_0 \; {\cal N}\left[\check{\phi}(\xi_1,\xi_2,z;q) \right], \label{bc:zero:phi}
\end{equation}
and
\begin{eqnarray}
&& (1-q) \check{\eta}(\xi_1,\xi_2;q)\nonumber\\ 
 &=& q \; c_0 \left\{ \check{\eta}(\xi_1,\xi_2;q)-\frac{1}{g}  \left[ {\sigma}_1\; \frac{\partial \check{\phi}(\xi_1,\xi_2,z;q)}{\partial \xi_1} + {\sigma}_2 \; \frac{\partial \check{\phi}(\xi_1,\xi_2,z;q)}{\partial \xi_2}  - \check{f} \right]\right\}, \hspace{0.75cm} \label{bc:eta:zero}
\end{eqnarray}
where
\[    \check{f} = \frac{1}{2} \hat{\nabla} \check{\phi}(\xi_1,\xi_2,z;q) \cdot  \hat{\nabla} \check{\phi}(\xi_1,\xi_2,z;q). \]
Besides, at the bottom, it holds
\begin{equation}
\frac{\partial \check{\phi}(\xi_1,\xi_2,z;q)}{\partial z} = 0, \;\;\;  \mbox{ as $z \rightarrow -\infty$}.  \label{bc:zero:bottom}
\end{equation}

Thus, using the property ${\cal L }\left( 0 \right) =0$ of the auxiliary linear operator (\ref{def:L}),  we have when $q=0$ the following relationships
\begin{equation}
\check{\phi}(\xi_1,\xi_2,z;0) = \phi_0(\xi_1,\xi_2,z),  \label{phi:q=0}
\end{equation}
and
\begin{equation}
\check{\eta}(\xi_1,\xi_2;0) = 0,  \label{eta:q=0}
\end{equation}
which provide us the initial approximations of the potential function $\phi(\xi_1,\xi_2,z)$ and the free surface $\eta(\xi_1,\xi_2)$.   Since $c_0 \neq 0$, when $q=1$,  Eqs. (\ref{geq:zero}) to (\ref {bc:zero:bottom}) are equivalent to the original equations (\ref{geq:phi}),(\ref{bc:eta}), (\ref{bc:phi}) and (\ref{bc:phi:bottom}), respectively, provided 
\begin{equation}
\check{\phi}(\xi_1,\xi_2,z;1) = \phi(\xi_1,\xi_2,z), \; \check{\eta}(\xi_1,\xi_2;1) = \eta(\xi_1,\xi_2).  \label{phi:q=1}
\end{equation}
So, as the embedding parameter $q\in[0,1]$ increases from 0 to 1,  $\check{\phi}(\xi_1,\xi_2,z;q)$ deforms (or varies) {\em continuously} from the initial approximation $\phi_0(\xi_1,\xi_2,z)$ to the unknown potential function $\phi(\xi_1,\xi_2,z)$, so does $\check{\eta}(\xi_1,\xi_2;q)$ from 0 to the unknown wave profile $\eta(\xi_1,\xi_2)$, respectively.   Mathematically speaking,  Eqs.~(\ref{geq:zero}) to (\ref {bc:zero:bottom})   define two  homotopies:
\begin{eqnarray}
\check{\phi}(\xi_1,\xi_2,z;q) & :=  &  \phi_0(\xi_1,\xi_2,z) \sim \phi(\xi_1,\xi_2,z), \nonumber \\
\check{\eta}(\xi_1,\xi_2;q) & := & 0 \sim \eta(\xi_1,\xi_2), \nonumber 
\end{eqnarray}
This is exactly the reason why   Eqs.~(\ref{geq:zero}) to (\ref {bc:zero:bottom})  are called {\em the zeroth-order deformation equations}.  

Note that we have freedom to choose the value of the convergence-control parameter $c_0$.    Assuming that  $c_0$ is so properly chosen that the Taylor series 
\begin{eqnarray}
\check{\phi}(\xi_1,\xi_2,z;q) &=& \phi_0(\xi_1,\xi_2,z) +\sum_{n=1}^{+\infty} \phi_n(\xi_1,\xi_2,z) \; q^n, \label{series:phi:q}\\
\check{\eta}(\xi_1,\xi_2;q) &=&  \sum_{n=1}^{+\infty} \eta_n(\xi_1,\xi_2) \; q^n,  \label{series:eta:q}
\end{eqnarray}
exist and converge at $q=1$,  then we have due to (\ref {phi:q=1}) the homotopy-series solution
\begin{eqnarray}
\phi(\xi_1,\xi_2,z) &=& \phi_0(\xi_1,\xi_2,z) +\sum_{n=1}^{+\infty} \phi_n(\xi_1,\xi_2,z), \label{series:phi}\\
\eta(\xi_1,\xi_2) &=&  \sum_{n=1}^{+\infty} \eta_n(\xi_1,\xi_2) ,  \label{series:eta}
\end{eqnarray}
where
\begin{eqnarray}
\phi_n(\xi_1,\xi_2,z) &=& \left. \frac{1}{n!} \frac{\partial^n \check{\phi}(\xi_1,\xi_2,z;q)}{\partial q^n}\right|_{q=0}, \nonumber\\
\eta_n(\xi_1,\xi_2) &=&  \left. \frac{1}{n!} \frac{\partial^n \check{\eta}(\xi_1,\xi_2;q)}{\partial q^n}\right|_{q=0}, \nonumber
\end{eqnarray}
are called homotopy-derivatives.   In the above formulas, the relationships  (\ref{phi:q=0}) and  (\ref{eta:q=0}) are used.    At the $m$th-order of approximations, we have
\begin{eqnarray}
\phi (\xi_1,\xi_2,z) &\approx & \phi_0 (\xi_1,\xi_2,z) + \sum_{n=1}^{m} \phi_n(\xi_1,\xi_2,z), \nonumber \\
\eta(\xi_1,\xi_2)    &\approx & \sum_{n=1}^{m} \eta_n(\xi_1,\xi_2).\nonumber
\end{eqnarray}

The equations for $\phi_n(\xi_1,\xi_2,z)$ and $\eta_n(\xi_1,\xi_2)$ can be  derived directly from the zeroth-order deformation equations  (\ref{geq:zero}) to (\ref {bc:zero:bottom}).  Substituting the series (\ref{series:phi:q}) into the governing equation (\ref{geq:zero}) and the boundary condition (\ref{bc:zero:bottom}) at bottom,  and equating the like-power of the embedding parameter $q$, we have 
\begin{equation}
\hat{\nabla} \; \phi_m(\xi_1,\xi_2,z) = 0, \;\;   \label{geq:phi:m}
\end{equation}
subject to the boundary condition at bottom
\begin{equation}
\frac{\partial \phi_m(\xi_1,\xi_2,z)}{\partial z} = 0, \;\; \mbox{as $z\rightarrow -\infty$}, \label{bc:phi:bottom:m}
\end{equation}
where $m\geq 1$.   It should be emphasized that  the two boundary conditions (\ref{bc:zero:phi}) and (\ref{bc:eta:zero}) are satisfied on the unknown boundary $z = \check{\eta}(\xi_1,\xi_2;q)$, which itself is now dependent upon the embedding parameter $q$, too.  So, it is relatively more complicated to deduce the corresponding equations.   Briefly speaking,  substituting the series (\ref{series:phi:q}) and (\ref {series:eta:q})  into the boundary condition (\ref{bc:zero:phi}) and (\ref{bc:eta:zero}) with  $z=\check\eta(\xi_1,\xi_2;q)$, then equating the like-power of $q$, we have two linear boundary conditions on $z = 0$:
\begin{eqnarray}
 \bar{\cal L} \left( \phi_m \right) &=& c_0 \; \Delta_{m-1}^\phi +\chi_m\; S_{m-1} -\bar{S}_m, \;\; m\geq 1,  \label{bc:phi:m}
\end{eqnarray}
and   
 \begin{eqnarray}
 \eta_m(\xi_1,\xi_2) &=&  c_0 \; \Delta_{m-1}^\eta  + \chi_m  \; \eta_{m-1}, \label{bc:eta:m}
 \end{eqnarray}
 where
 \begin{equation}
 \Delta_{m-1}^\eta = \eta_{m-1} -\frac{1}{g} \left[  \left(\sigma_{1} \; \bar\phi_{m-1}^{1,0}
 + \sigma_{2}  \; \bar\phi_{m-1}^{0,1} \right) -\Gamma_{m-1,0}\right] \label{def:Delta:eta}
 \end{equation}
 and 
\begin{equation}
\bar{\cal L} \left( \phi_m \right) = \left. \left( \bar\sigma_1^2 \frac{\partial^2 \phi_m}{\partial \xi_1^2}
 +2\; \bar\sigma_1\bar\sigma_2\; \frac{\partial^2\phi_m}{\partial\xi_1\partial\xi_2}
 + \bar\sigma_2^2 \frac{\partial^2 \phi_m}{\partial \xi_2^2}+ g \; \frac{\partial \phi_m}{\partial z}\right)\right|_{z=0} . 
 \label{def:L:z=0} 
\end{equation}
The detailed derivation of the above equations and the definitions of $\Delta^\phi_{m-1}$, $S_{m-1}$, $\bar{S}_m$, $\chi_m$, $\Gamma_{m-1,0}$, $\bar\phi_{m-1}^{1,0}$, $\bar\phi_{m-1}^{0,1}$  are given in Appendix~A.   Note that, the sub-problems for $\phi_{m}$ and $\eta_{m}$ are not only {\em linear} but also {\em decoupled}:  given $\phi_{m-1}$ and $\eta_{m-1}$, it is straightforward to get $\eta_{m}$ directly,  and then $\phi_{m}$ is obtained by solving the linear Laplace equation (\ref{geq:phi:m}) with two linear boundary conditions (\ref {bc:phi:bottom:m}) and (\ref{bc:phi:m}).    Thus, the high-order deformation equations can be easily solved by means of the symbolic computation software.   

Liao \cite{LiaoBook2003}  proved in general that the homotopy-series solutions satisfy the original nonlinear equations as long as they are convergent.  Similarly, we have such a theorem:\\ \\
{\bf Convergence Theorem} {\em The homotopy-series solution (\ref{series:phi}) and (\ref{series:eta}) satisfy the original governing equation (\ref{geq:phi})  and the boundary conditions (\ref{bc:eta}),  (\ref{bc:phi}) and  (\ref{bc:phi:bottom}),  provided that }
\begin{equation}
\sum_{m=0}^{+\infty} \Delta_m^\phi = 0, \hspace{1.0cm}  \sum_{m=0}^{+\infty} \Delta_m^\eta = 0, \label{def:criterion:convergence}
\end{equation}
{\em where $\Delta_m^\phi, \Delta_m^\eta$ are defined by (\ref{def:Delta:phi}) and (\ref{def:Delta:eta}), respectively}.  \\ \\
 A  mathematical proof of the above convergence theorem is given briefly  in Appendix~B.   Because the terms $\Delta_m^\phi$ and $\Delta_m^\eta$ are by-products in solving high-order deformation equations, the above theorem provides us a convenient way to check the convergence and accuracy of the homotopy-series solution.  For this reason, we define the residual error squares
 \begin{eqnarray}
 {\cal E}^\phi_m  &=&  \frac{1}{\pi^2}\int_{0}^{\pi}\int_0^{\pi}\left( \sum_{n=0}^{m}\Delta^\phi_n\right)^2 d \xi_1 \; d\xi_2, \label{def:E:phi}\\
 {\cal E}^\eta_m  &=&  \frac{1}{\pi^2}\int_{0}^{\pi}\int_0^{\pi}\left( \sum_{n=0}^{m}\Delta^\eta_n\right)^2 d \xi_1 \; d\xi_2, \label{def:E:eta}
 \end{eqnarray} 
 for the $m$th-order approximations of $\phi$ and $\eta$.     According to this convergence theorem,  the homotopy-series solution (\ref{series:phi}) and (\ref{series:eta})  satisfy the original equation and all boundary conditions if ${\cal E}^\phi_m\rightarrow 0$ and ${\cal E}_m^{\eta}\rightarrow 0$  as $m\rightarrow +\infty$.   Besides, the values of ${\cal E}_m^\phi$ and ${\cal E}_m^\eta$ indicate the accuracy of the $m$th-order approximation of $\phi$ and $\eta$, respectively.

Note that  the auxiliary linear operator (\ref{def:L}) has the property
\begin{eqnarray}
{\cal L}\Psi_{m,n}  &=&\left[  g | m{\bf k}_1 + n {\bf k}_2 |-(m \bar\sigma_1+n \bar\sigma_2)^2 \right]\Psi_{m,n} ,\hspace{1.0cm} \label{def:L:property}
\end{eqnarray}
where $\Psi_{m,n}$ is defined by (\ref{def:psi}).
Thus, mathematically speaking,  $\cal L$ has an infinite number of eigenfunctions $\Psi_{m,n}$  with the corresponding eigenvalue
\begin{equation}
\lambda_{m,n} =    g | m{\bf k}_1 + n {\bf k}_2 |-\left(m\bar\sigma_2+n\bar\sigma_2 \right)^2. \label{def:eigenvalue}
\end{equation}
In short, 
\[    {\cal L}  \left( \Psi_{m,n} \right)  =  \lambda_{m,n} \; \Psi_{m,n}.   \]
Besides,  $\Psi_{m,n}$ defined by (\ref{def:psi}) automatically satisfies the governing equation (\ref{geq:phi}) and the boundary condition (\ref{bc:phi:bottom}) at the bottom.    Therefore, the  inverse operator ${\cal L}^{-1}$ is defined by 
\begin{eqnarray}
{\cal L}^{-1} \left(  \Psi_{m,n}   \right) 
= \frac{ \Psi_{m,n}} {\lambda_{m,n}} ,  \hspace{1.0cm}  \lambda_{m,n}\neq 0 . \label{def:L:inverse}
\end{eqnarray}
Note that the inverse operator ${\cal L}^{-1}$ has definition {\em only} for non-zero eigenvalue $\lambda_{m,n}$.      When $\lambda_{m,n}=0$, we have    
\begin{equation}
g  | m{\bf k}_1 + n {\bf k}_2 | = ( m\bar\sigma_1 + n\bar\sigma_2 )^2, \label{resonance}
\end{equation}
corresponding to the criterion of the so-called ``wave resonance'' mentioned by Phillips and Longuet-Higgins.    Thus,  the number of eigenfunctions with zero eigenvalue, denoted by $N_\lambda$,  is the key for solving the problem.  

When $n=0$,  it holds 
\[ \lambda_{m,0} = k_1 \left( |m|-m^2  \right),  \]
which equals to zero only when $|m|=1$ (note that $m=n=0$ corresponds to $\phi=0$ and thus is not considered here).  Similarly, when $m=0$, the eigenvalue $\lambda_{0,n}$  equals to zero only when $|n| =  1$.   So, there exist at least  two eigenfunctions   $ \Psi_{1,0} = e^{k_1 z} \sin \xi_1 $ and $ \Psi_{0,1} =  e^{k_2 z} \sin \xi_2$ whose eigenvalues are  zero,  i.e. 
\begin{equation}
{\cal L} \left(C_1 \;  e^{k_1 z} \sin \xi_1 + C_2 \;  e^{k_2 z}  \sin \xi_2\right)  = 0 \label{L:property:2}
\end{equation}  
for any constants $C_1$ and $C_2$ independent of $\xi_1,\xi_2$ and $z$.  Therefore, it holds  $N_\lambda \geq 2$ in case of  two primary waves.    As mentioned by Phillips \cite{Phillips1960JFM} and Longuet-Higgins \cite{LonguetHiggins1962JFM},  the  criterion  (\ref{resonance}) of wave resonance can be satisfied for some special wave numbers and angular frequencies.  Thus, 
when  (\ref{resonance}) is satisfied in case of $m=m'$ and $n=n'$,   where $m'$ and $n'$ are integers with $m'^2 + n'^2 \neq 1$,  there exist three eigenfunctions   $\Psi_{1,0}=  e^{k_1 z} \sin \xi_1$, $\Psi_{0,1}= e^{k_2 z} \sin \xi_2$ and 
\[  \Psi_{m',n'} =  e^{| m'{\bf k}_1 + n' {\bf k}_2 | z} \sin(m'\xi_1+n'\xi_2) \] 
 whose eigenvalues are zero, i.e.
 \begin{equation}   
  {\cal L} \left[ C_1 \;  e^{k_1 z} \sin \xi_1 + C_2 \;  e^{k_2 z}  \sin \xi_2 +C_3 \; e^{| m'{\bf k}_1 + n' {\bf k}_2 | z} \sin(m'\xi_1+n'\xi_2) \right]  =0  \label{L:property:3}
  \end{equation}
  for any constants $C_1, C_2$ and $C_3$ independent of $\xi_1, \xi_2$ and $z$.  
     Without loss of generality, Longuet-Higgins  \cite{LonguetHiggins1962JFM} discussed a special case  $m'=2$ and $n'=-1$, corresponding to the eigenfunction \[\Psi_{2,-1}=\exp(|2{\bf k}_1-{\bf k}_2|z) \sin(2\xi_1-\xi_2).\]  In \S 3 of this article,  it implies $m'=2$ and $n'=-1$ when $N_\lambda =3$ in case of two primary waves, if not explicitly mentioned.   
    
  According to the definitions (\ref{def:L}) and (\ref{def:L:z=0}),  it holds 
  \[   \bar{\cal L} \phi = \left. \left( {\cal L} \phi \right)  \right|_{z=0}.  \]
  Thus,  
\[   \bar{\cal L}\left( \Psi_{m,n}  \right) = \left. \lambda_{m,n} \; \Psi_{m,n}\right|_{z=0}  = \lambda_{m,n} \sin(m\xi_1+n\xi_2), \]
which gives the definition of  the linear inverse operator
\begin{equation}
\bar{\cal L}^{-1} \left[ \sin(m\xi_1+n\xi_2) \right] = \frac{\Psi_{m,n}}{\lambda_{m,n}}, \;\;\;  \lambda_{m,n} \neq 0.  \label{def:L:inverse:z=0}
\end{equation}
Using this inverse operator, it is easy to solve the linear Laplace equation (\ref{geq:phi:m}) with two linear boundary conditions (\ref {bc:phi:bottom:m}) and (\ref{bc:phi:m}), as illustrated  below.   Here, we emphasize that the above inverse operator has definition only for non-zero eigenvalue $\lambda_{m,n} \neq 0$.

\subsection{In case of $N_\lambda = 2$: non-resonant waves}

In this case, there are only two eigenfunctions  
\[ \Psi_{1,0}= \exp(k_1 z)   \sin \xi_1, \;\;\; \Psi_{0,1}= \exp(k_2 z)   \sin \xi_2 \] 
whose eigenvalues are zero, i.e. $\lambda_{1,0}=\lambda_{0,1}=0$.   Using these two eigenfunctions and according to the linear wave theory,   we construct  the initial approximation of the potential function
\begin{equation}
\phi_0(\xi_1,\xi_2,z)  = A_0 \; \sqrt{\frac{g}{k_1}}  \; \Psi_{1,0} +B_0 \; \sqrt{\frac{g}{k_2}} \; \Psi_{0,1},   \label{def:phi[0]}
\end{equation}
where  $A_0$ and $B_0$ are unknown constants.  

Using formulas mentioned above or in Appendix A,  the  corresponding first-order deformation equation about the potential function $\phi_1(\xi_1,\xi_2,z)$ reads
\begin{equation}
\hat{\nabla}\phi_1(\xi_1,\xi_2,z) = 0,   \label{geq:phi[1]}
\end{equation}
subject to the boundary condition on $z=0$:
\begin{eqnarray}
\bar{\cal L}\left(  \phi_1  \right) 
&=& b_{1}^{1,0} \sin(\xi_1)  +  b_{1}^{0,1} \sin(\xi_2) + b_{1}^{1,1} \sin(\xi_1+\xi_2) + d_{1}^{1,1} \sin(\xi_1-\xi_2)\nonumber\\
&+&  b_{1}^{2,1} \sin(2\xi_1+\xi_2) +  d_{1}^{2,1} \sin(2\xi_1-\xi_2)\nonumber\\
&+&  b_{1}^{1,2} \sin(\xi_1+2\xi_2)+ d_{1}^{1,2} \sin(\xi_1-2\xi_2),
\end{eqnarray}
and  the boundary condition on the bottom:
\begin{equation}
\left.  \frac{\partial \phi_1}{\partial z}\right|_{z=0}  =  0,   \label{bc:phi[1]:bottom}
\end{equation}
where $\bar{\cal L}$ is defined by (\ref{def:L:z=0}), and  $b_1^{i,j}, b_1^{i,j}$ are constants.  Especially,   we have 
\begin{eqnarray}
b_1^{1,0} &=& c_0 A_{0}\sqrt{ \frac{g   }{k_1} }  \left[ g k_1-\sigma_{1}^2 +g k_1 (A_0 k_1)^2+ 2 g k_1 (B_0 k_2)^2 +\frac{g  B_0^2 k_1^2 k_2 }{2} \sin^2(\alpha_1-\alpha_2)  \right],\nonumber\;\;\;\\
b_1^{0,1} &=& c_0 B_{0}\sqrt{ \frac{g   }{k_2} }  \left[ g k_2-\sigma_{2}^2 +g k_2 (B_0 k_2)^2+ 2 g k_2 (A_0 k_1)^2 +\frac{g  A_0^2 k_2^2 k_1 }{2} \sin^2(\alpha_1-\alpha_2)  \right]. \nonumber
\end{eqnarray}
Since $\lambda_{1,0}=\lambda_{0,1}=0$, according to (\ref{def:L:inverse:z=0}),  it must hold 
\[    b_1^{1,0} = b_1^{0,1}=0 \]
so as to avoid the so-called  ``secular''  terms $\xi_1 \sin\xi_1$ and $\xi_2\sin\xi_2$.   This  provides us  the following  algebraic equations
\begin{eqnarray}
 (A_0 k_1)^2+ \left[2 +\frac{  k_1 }{2 k_2} \sin^2(\alpha_1-\alpha_2)\right] (B_0 k_2)^2  =\frac{\sigma_{1}^2}{g k_1} -1,\\
\left[2 +\frac{  k_2 }{2 k_1} \sin^2(\alpha_1-\alpha_2)\right] (A_0 k_1)^2 +   (B_0 k_2)^2 =\frac{\sigma_{2}^2}{g k_2} -1,
\end{eqnarray}
whose solutions are 
\begin{eqnarray}
A_0 &=& \pm \left(\frac{\varepsilon}{k_1}\right)\sqrt{\left(\frac{\sigma_2^2}{g k_2}-1\right) \left[2 +\frac{  k_1 }{2 k_2} \sin^2(\alpha_1-\alpha_2)\right]- \left(\frac{\sigma_1^2}{g k_1}-1\right) } \; \; , \label{def:A[0]}\\
B_0 &=& \pm \left(\frac{\varepsilon}{k_2}\right) \sqrt{\left(\frac{\sigma_1^2}{g k_1}-1\right) \left[2 +\frac{  k_2 }{2 k_1} \sin^2(\alpha_1-\alpha_2)\right]-\left(\frac{\sigma_2^2}{g k_2}-1\right)}\;\; ,\label{def:B[0]}
\end{eqnarray}
where
\[   \varepsilon  = \left( \left[2 +\frac{  k_1 }{2 k_2} \sin^2(\alpha_1-\alpha_2)\right] \left[2 +\frac{  k_2 }{2 k_1} \sin^2(\alpha_1-\alpha_2)\right]-1  \right)^{-1/2}.   \]
Note that $A_0$ and $B_0$ have multiple values: they can be either positive or negative.  

Then,  by means of  the linear inverse operator $\bar{\cal L}^{-1}$ defined by (\ref {def:L:inverse:z=0}),  the common solution of $\phi_1(\xi_1,\xi_2,z)$   reads
\begin{eqnarray}
\phi_1 &=& A_1  \sqrt{\frac{g}{k_1}}   \; \Psi_{1,0} + B_1 \sqrt{\frac{g}{k_2}}   \;  \Psi_{0,1} 
+   {b}_1^{1,1}\left(\frac{\Psi_{1,1}}{\lambda_{1,1}} \right) + {d}_1^{1,1} \left( \frac{\Psi_{1,-1}}{\lambda_{1,-1}}\right)\nonumber\\
&+&  {b}_1^{2,1} \left(\frac{\Psi_{2,1}}{\lambda_{2,1}} \right) + {d}_1^{2,1} \left(\frac{\Psi_{2,-1}}{\lambda_{2,-1}} \right)
   +     {b}_1^{1,2} \left( \frac{\Psi_{1,2}}{\lambda_{1,2}} \right)+ {d}_1^{1,2} \left(\frac{\Psi_{1,-2}}{\lambda_{1,-2}} \right), \label{def:phi[1]}
\end{eqnarray}
where $A_1$ and $B_1$ are unknown coefficients,  the eigenfunction $\Psi_{m,n}$ and eigenvalue $\lambda_{m,n}$ are defined by (\ref{def:psi}) and (\ref{def:eigenvalue}), respectively.    In other words, $\phi_1$ is a sum ( or linear combination) of  eigenfunctions.   
Note that $\phi_1$ automatically satisfies the Laplace equation (\ref{geq:phi:m}) and the bottom condition (\ref {bc:phi:bottom:m}) for {\em any} constants $A_1$ and $B_1$.    On the other hand, given the initial guess $\phi_0$,  it is straightforward to calculate $\eta_1(\xi_1,\xi_2)$  directly by means of the formula (\ref{bc:eta:m}). 

The above approach has general meaning.  In a similar way,  we can obtain $\eta_{m}(\xi_1,\xi_2)$ and  $\phi_m(\xi_1,\xi_2,z)$,  successively, in the order $m=1,2,3,$ and so on.    Note that, the two unknown coefficients $A_m$ and $B_m$ ($m\geq 1$)  can be determined exactly in the same way like $A_0$ and $B_0$ by means of avoiding the ``secular'' terms $\xi_1 \sin\xi_1$ and $\xi_2\sin\xi_2$.    Note that only fundamental operations are needed  in the above approach so that it is convenient to use symbolic computations to get high-order approximations.

Without loss of generality, let us consider  here such a special case of the two primary waves that 
\begin{equation}
 \frac{\sigma_1}{\sqrt{g k_1}} = \frac{\sigma_2}{\sqrt{g k_2}} =1.0003,  \;\; \alpha_1=0, \;\; \alpha_2=\frac{\pi}{36},\;\; k_2 = \frac{\pi}{5},
 \label{def:case:A}
 \end{equation}
with different ratios of $k_1/k_2$.   Here,  the number 1.0003  is chosen so that the perturbation theory is valid with high accuracy and thus we can compare our results with those given by Phillips \cite{Phillips1960JFM} and Longuet-Higgins \cite{LonguetHiggins1962JFM},  who    suggested that the wave resonance (with amplitude growing in time) occurs if the criterion (\ref{resonance}) is satisfied, i.e. 
\[     \frac{k_2}{k_1} \approx  0.8925, \;\;\;   \frac{\sigma_1}{\sigma_2} = \sqrt{\frac{k_1}{k_2}}\approx 1.0585       \] 
in the current case.   As mentioned before,  $N_\lambda = 3$ when the above criterion  is satisfied.   To avoid this, let us first consider here the non-resonant waves with different  wave number $k_1$   except  $k_2/k_1 = 0.8925$.

\begin{figure}
\setcaptionwidth{5in} \centering
\includegraphics[scale=0.4]{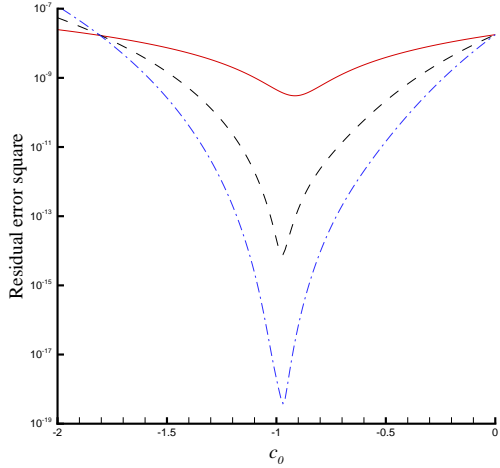}
\caption{Residual error square versus the convergence control parameter $c_0$ in case of (\ref{def:case:A}) with $k_2/k_1=1$.}\label{figure:error}
\end{figure}

\begin{table}[htdp]
\caption{Residual error square of two nonlinear boundary conditions in case of (\ref{def:case:A}) with $k_2/k_1=1$}
\begin{center}
\begin{tabular}{|c|c|c|}\hline\hline
$m$ & ${\cal E}_m^\phi$ & ${\cal E}_m^\eta$ \\  \hline
1	&	1.9 $\times 10^{-8}$ 	&	5.1  $\times 10^{-4}$ \\
3	&	3.5 $\times 10^{-12}$    &	1.2 $\times 10^{-9}$	\\
5	&	2.0 $\times 10^{-16}$	&	4.3 $\times 10^{-14}$	\\
8	&	2.8	$\times 10^{-22}$	&	3.8$\times 10^{-20}$   \\
10	&	4.7	$\times 10^{-26}$	&   6.4	$\times 10^{-24}$  \\
\hline\hline
\end{tabular}
\end{center}
\label{Table:N=2:error}
\end{table}

First of all, we consider a special case $k_2/k_1=1$.   Without loss of generality, we take negative values of $A_0$ and $B_0$ given by (\ref{def:A[0]}) and (\ref{def:B[0]}), respectively.       To choose an optimal value of the  convergence-control parameter $c_0$  so that the series solution of $\phi$ and $\eta$ converge quickly,   we plot the curves of the residual error squares ${\cal E}_m^\phi$ and ${\cal E}_m^\eta$  versus $c_0$, as shown in Fig.~\ref{figure:error}.   When  ${\cal E}_m^\phi$ and ${\cal E}_m^\eta$ contain the unknown convergence-control parameter $c_0$, the related integrals are rather time-consuming.   To avoid this,  a discrete technique suggested by Liao \cite{LIAO2010-CNSNS-B}  is used.    It is found that the residual error square  ${\cal E}_m^\phi$  decreases  for  $-1.8 \leq c_0 <0 $ and the optimal value of $c_0$ is close to -1, as  shown in Fig.~\ref{figure:error}.  Therefore, we choose $c_0 = -1$, and the corresponding residual error squares of the two boundary conditions decrease rather quickly to the level $10^{-24}$ at the 10th-order approximation, as listed in Table~\ref{Table:N=2:error}.    According to the Convergence Theorem mentioned above, the corresponding homotopy-series (\ref{series:phi}) and (\ref{series:eta}) are  the solution of the problem.

\begin{figure}
\setcaptionwidth{5in} \centering
\includegraphics[scale=0.4]{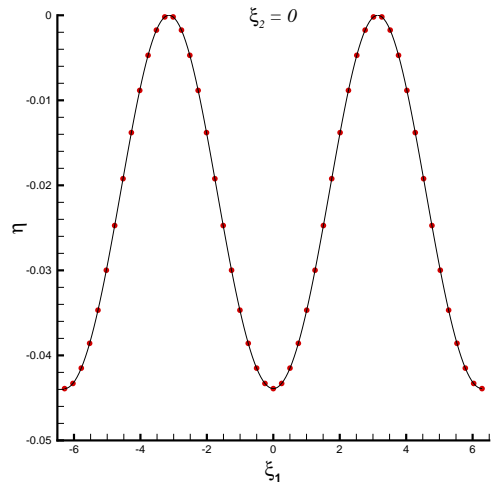}
\caption{Wave profile at $\xi_2=0$  in case of (\ref{def:case:A}) with $k_2/k_1=1$. Symbols: 3rd-order approximation; Solid line: 10th-order approximation.}\label{figure:Eta:k1=k2}
\end{figure}

Let $a_{1,0}, a_{0,1}$ and $a_{2,-1}$ denote the amplitudes of wave  components $\cos\xi_1$, $\cos\xi_2$ and $\cos(2\xi_1-\xi_2)$, respectively.   Obviously, $a_{1,0}=a_{0,1}$ in case of  $k_1=k_2$.   As shown in Table~\ref{Table:N=2:wave-amplitude},  each wave component converges rather quickly, which agree well (see Table~\ref {Table:N=2:wave-amplitude:HP}) with those obtained by the homotopy-Pad\'{e} method \cite{Liao2007-SAM, LiaoBook2003}, a kind of acceleration technique developed in the frame of the HAM.   Besides, the analytic approximations of wave profile at $\xi_2=0$ also converge quickly, as shown in Fig.~\ref{figure:Eta:k1=k2}.   All of these indicate the validity of the analytic approach based on the HAM.

\begin{table}[htdp]
\caption{Analytic approximations of wave amplitude components in case of (\ref{def:case:A}) with $k_2/k_1=1$ }
\begin{center}
\begin{tabular}{|c|c|c|} \hline\hline
Order of appr.  &	$a_{1,0}, a_{0,1}$ 	&	$a_{2,-1}$   \\  \hline
1	&	-0.022502	&	0	\\
2	&	-0.022846	&	0.00059739	\\
3	&	-0.022814	&	0.00057198	\\	
4	&	\underline{-0.022816}	&	0.00057208	\\
6	&	-0.022816	&	\underline{0.00057226}	\\
8	&	-0.022816	&	0.00057226	\\
10	&	-0.022816	&	0.00057226	\\
\hline\hline
\end{tabular}
\end{center}
\label{Table:N=2:wave-amplitude}
\end{table}

\begin{table}[htdp]
\caption{Analytical approximations of wave amplitude components given by the $[m,m]$ homotopy-Pad\'{e}  method in case of   (\ref{def:case:A}) with $k_2/k_1=1$}
\begin{center}
\begin{tabular}{|c|c|c|}\hline\hline 
$m$	&	$a_{1,0}, a_{0,1}$	&	$a_{2,1}$ \\  \hline
2	&	\underline{-0.022816}	&	0.00057211	\\
3	&	-0.022816	&	\underline{0.00057226}	\\
4	&	-0.022816	&	0.00057226	\\
5	&	-0.022816	&	0.00057226	\\
\hline\hline
\end{tabular}
\end{center}
\label{Table:N=2:wave-amplitude:HP}
\end{table}

Similarly, we can get convergent series solutions in case of (\ref{def:case:A}) with different ratio of $k_2/k_1$, as shown in Table~\ref{Table:wave-amplitude:all}.  Note that, as the ratio of $k_2/k_1$ decreases,  the wave amplitude of the component $\cos(2\xi_1-\xi_2)$ first increases monotonically to the maximum value at $k_2/k_1=0.8925$ (the corresponding result is given by the approach described in \S 3.2), and then decreases monotonically.   Note that the variation of $a_{1,0}$ and $a_{0,1}$ is not monotonic.   It must be emphasized that  $|a_{1,0}|$ and $|a_{0,1}|$ is much {\em larger}  than $|a_{2,1}|$ for {\em all} values of $k_2/k_1$, as shown in Table~ \ref{Table:wave-amplitude:all}.  We will discuss this interesting phenomena later in details. 
	
	\begin{table}[htdp]
\caption{Wave amplitude components in case of (\ref{def:case:A}) with different values of $k_2/k_1$}
\begin{center}
\begin{tabular}{|c|c|c|c|} \hline\hline
$k_2/k_1$	&	$a_{1,0}$	&	$a_{0,1}$	&	$|a_{2,1}|$	\\ \hline
1.00	&	-0.0228	&	-0.0228	&	0.00057	\\
0.95	&	-0.0229	&	-0.0218	&	0.00084	\\
0.93	&	-0.0230	&	-0.0215	&	0.00116	\\
0.92	&	-0.0231	&	-0.0214	&	0.00150	\\
0.91	&	-0.0231	&	-0.0215	&	0.00219	\\
0.905	&	-0.0232	&	-0.0216	&	0.00290	\\
0.90	&	-0.023	&	-0.022	&	0.00433	\\
\underline{0.8925}	&	\underline{-0.0205}	&	\underline{-0.0232}	&	\underline{0.00898} 	\\
0.88	&	-0.0235	&	-0.0179	&	0.00307	\\	
0.86	&	-0.0221	&	-0.0189	&	0.00120	\\
0.85	&	-0.0226	&	-0.0187	&	0.00087	\\
0.83	&	-0.0229	&	-0.0182	&	0.00054	\\
0.80	&	-0.0232	&	-0.0172	&	0.00033	\\
0.70	&	-0.0239	&	-0.0121	&	0.00010	\\
\hline\hline
\end{tabular}
\end{center}
\label{Table:wave-amplitude:all}
\end{table}

\subsection{In case of $N_\lambda = 3$: resonant waves}

Now, let us further consider the case of (\ref{def:case:A}) with a special ratio  $k_2/k_1 = 0.8925$ for two primary traveling waves with wave resonance.    It was suggested first by Phillips \cite{Phillips1960JFM} and then confirmed by Longuet-Higgins \cite{LonguetHiggins1962JFM} that the so-called wave resonance occurs in this case so that the  amplitude of  wave component $\cos(2\xi_1-\xi_2)$ grows in time, i.e.  $a_{2,-1} = \bar{\alpha} \; t$ , where $\bar{\alpha} $ is a constant.  

In this case, the criterion (\ref{resonance}) is satisfied, i.e. 
\[     g|2{\bf k}_1-{\bf k}_2| = (2\bar\sigma_1-\bar\sigma_2)^2.  \]
According to  (\ref{def:case:A}), we have
\[   k_1 = 0.703998, \;\; k_2 = 0.628319,\;\; k_3= |2{\bf k}_1-{\bf k}_2| = 0.783981.   \]
Thus, according to (\ref{def:eigenvalue}),   we have an additional zero eigenvalue $\lambda_{2,-1} = 0$ with the corresponding eigenfunction 
\[   \Psi_{2,-1}  =  e^{|2{\bf k}_1-{\bf k}_2|z}  \sin(2\xi_1-\xi_2).  \] 
So, we have now {\em three} eigenfunctions $\Psi_{1,0}, \Psi_{0,1}$ and $\Psi_{2,-1}$ whose eigenvalues are zero, i.e.  $\lambda_{1,0}=0$,  $\lambda_{0,1}=0$ and $\lambda_{2,-1}=0$.   Here, it should be emphasized that the wave resonance condition given by Phillips \cite{Phillips1960JFM} is mathematically equivalent to that the eigenvalue of nonlinear-interaction wave is zero.   
Unfortunately, the new zero eigenvalue $\lambda_{2,-1}$  breaks down the approach mentioned in \S 3.1, because the term $\Psi_{2,-1}/\lambda_{2,-1}$ in (\ref{def:phi[1]}) becomes infinite: this is exactly the reason why Phillips \cite{Phillips1960JFM}  and Lenguet-Higgins \cite{LonguetHiggins1962JFM} suggested the existence of the so-called wave-resonance with amplitude growing in time, which, however, is physically impossible from the view-points of wave energy.  

 Can we avoid such kind of wave-resonance with amplitude growing in time?  
 
 Note that, in case of non-resonant waves investigated in \S 3.1,  the initial guess (\ref{def:phi[0]})  is a linear combination of the {\em two} eigenfunctions $\Psi_{1,0}$ and $\Psi_{0,1}$ whose eigenvalues $\lambda_{1,0}$ and $\lambda_{0,1}$ are zero.  In the current case of wave resonance, the only difference is that we have an additional eigenfunction $\Psi_{2,-1}$ whose eigenvalue $\lambda_{2,-1}$ is zero, too.   As mentioned in many other publications \cite{LiaoBook2003, liao99-nlm, liao99-jfm, liao02-jfm, liao03-jfm, Liao2006-SAM, Liao2007-SAM, Xu2010-PhysicFluids, Li2010-JMP},  the HAM provides us with great freedom to choose the initial guess.   With such kind of freedom, why not use these three eigenfunctions (with zero eigenvalue) to express the initial guess $\phi_0$?       In other words, we can express the initial guess $\phi_0$ by {\em all} eigenfunctions whose eigenvalues are zero, i.e. 
 \begin{eqnarray}
 \phi_0(\xi_1,\xi_2,z)  &=& \bar{A}_0 \; \sqrt{\frac{g}{k_1}}  \; \Psi_{1,0} +\bar{B}_0 \; \sqrt{\frac{g}{k_2}} \; \Psi_{0,1}   
+\bar{C}_0 \; \sqrt{\frac{g}{k_3}} \; \Psi_{2,-1},   \label{def:phi[0]:2}
 \end{eqnarray}
 where $\bar{A}_0, \bar{B}_0, \bar{C}_0$ are {\em unknown} constants independent of $\xi_1, \xi_2$ and $z$.  Similarly, substituting the above expression into  the deformation equations (\ref{geq:phi:m}) to (\ref{bc:phi:m}),   we have the same first-order deformation equation (\ref{geq:phi[1]}) with the same boundary condition (\ref{bc:phi[1]:bottom}) at bottom for $\phi_1$, but a more complicated  boundary condition on $z=0$, i.e.
 \begin{eqnarray}
\bar{\cal L}\left(  \phi_1  \right) 
&=& \bar{b}_{1}^{1,0} \sin(\xi_1)  +  \bar{b}_{1}^{0,1} \sin(\xi_2) + \bar{b}_{1}^{2,0} \sin(2\xi_1)    + \bar{b}_{1}^{3,0} \sin(3\xi_1)  \nonumber\\
&+& \bar{b}_{1}^{1,1} \sin(\xi_1+\xi_2)      + \bar{d}_{1}^{1,1} \sin(\xi_1-\xi_2)\nonumber\\
&+&  \bar{b}_{1}^{2,1} \sin(2\xi_1+\xi_2)   + \bar{d}_{1}^{2,1} \sin(2\xi_1-\xi_2)\nonumber\\
&+&  \bar{b}_{1}^{1,2} \sin(\xi_1+2\xi_2)   + \bar{d}_{1}^{1,2} \sin(\xi_1-2\xi_2) \nonumber\\
&+&   \bar{d}_{1}^{2,2} \sin(2\xi_1-2\xi_2) + \bar{d}_{1}^{2,3} \sin(2\xi_1-3\xi_2) \nonumber\\
&+&   \bar{d}_{1}^{3,1} \sin(3\xi_1-\xi_2)   + \bar{d}_{1}^{3,2} \sin(3\xi_1-2\xi_2) \nonumber\\
&+&   \bar{d}_{1}^{4,1} \sin(4\xi_1-\xi_2)   + \bar{d}_{1}^{4,2} \sin(4\xi_1-2\xi_2) + \bar{d}_{1}^{4,3} \sin(4\xi_1-3\xi_2)\nonumber\\
&+&   \bar{d}_{1}^{5,2} \sin(5\xi_1-2\xi_2) + \bar{d}_{1}^{6,3} \sin(6\xi_1-3\xi_2),\label{bc:phi[1]:2}
\end{eqnarray}  
where $\bar{b}_1^{m,n}, \bar{d}_1^{m,n}$ are constant  coefficients, and the linear operator $\bar{\cal L}$ is defined by (\ref{def:L:z=0}).  
Note that there exist now {\em three} zero eigenvalues, i.e.  $\lambda_{1,0}=0$,  $\lambda_{0,1}=0$ and $\lambda_{2,-1}=0$.    Therefore, according to the definition (\ref{def:L:inverse:z=0}) of the inverse operator $\bar{\cal L}^{-1}$,  {\em not only} the two coefficients $\bar{b}_{1}^{1,0} ,  \bar{b}_{1}^{0,1} $ {\em but also} the additional coefficient  $\bar{d}_{1}^{2,1} $  must be zero.   Enforcing  
\[ \bar{b}_{1}^{1,0} =0 ,  \;\;  \bar{b}_{1}^{0,1} =0 ,  \;\; \bar{d}_{1}^{2,1} = 0,\]
we obtain a set of nonlinear algebraic equations
\begin{equation}
\left\{ 
\begin{array}{lcc}
12.7576 \bar{A}_0^2+20.3675 \bar{B}_0^2 + 31.6768 \bar{C}_0^2 +25.6718\bar{B}_0 \bar{C}_0 &=& 0.0154469,\\
24.1456 \bar{A}_0^2+9.6004 \bar{B}_0^2 + 30.0398 \bar{C}_0^2 +14.9558\bar{A}_0^2 \bar{C}_0/\bar{B}_0 &=& 0.014593,\\
26.9621 \bar{A}_0^2+21.6116 \bar{B}_0^2 + 16.6956 \bar{C}_0^2 +10.7158\bar{A}_0^2 \bar{B}_0/\bar{C}_0 &=& 0.0163008
\end{array}
\right.  \label{eq:ABC}
\end{equation}
for the special case mentioned above.  The set of these nonlinear algebraic equations has four complex and  twelve real solutions.  Because the complex solutions have no physical meanings,  we list  only its twelve real roots in Table~\ref{Table:roots}.  It is found that the twelve roots fall into three groups, and different groups give different solutions, as shown later.   After solving this set of nonlinear algebraic equations, the initial guess 
$\phi_0$ is known and therefore it is straightforward to get $\eta_1$ directly by means of (\ref{bc:eta:m}).   More importantly,  on the right-hand side of Eq. (\ref{bc:phi[1]:2}),  the terms $\sin\xi_1$,  $\sin\xi_2$ and especially $\sin(2\xi_1-\xi_2)$  disappear now.  Then, using the inverse operator (\ref{def:L:inverse:z=0}),  it is straightforward to get the common solution of the first-order approximation  
 \begin{eqnarray}
 \phi_1  
&=&  \bar{A}_1 \sqrt{\frac{g}{k_1}}\; \Psi_{1,0}  + \bar{B}_1 \sqrt{\frac{g}{k_2}}\; \Psi_{0,1}   +  \bar{C}_1 \sqrt{\frac{g}{k_3}}\; \Psi_{2,-1} \nonumber\\
& +& \bar{b}_{1}^{2,0} \left( \frac{\Psi_{2,0}}{\lambda_{2,0}} \right)   + \bar{b}_{1}^{3,0} \left( \frac{\Psi_{3,0}}{\lambda_{3,0}} \right)   
    + \bar{b}_{1}^{1,1} \left( \frac{\Psi_{1,1}}{\lambda_{1,1}} \right)  + \bar{d}_{1}^{1,1} \left( \frac{\Psi_{1,-1}}{\lambda_{1,-1}} \right) \nonumber\\
&+&  \bar{b}_{1}^{2,1} \left( \frac{\Psi_{2,1}}{\lambda_{2,1}} \right)  
   +  \bar{b}_{1}^{1,2} \left( \frac{\Psi_{1,2}}{\lambda_{1,2}} \right) + \bar{d}_{1}^{1,2} \left( \frac{\Psi_{1,-2}}{\lambda_{1,-2}} \right)  + \bar{d}_{1}^{2,2} \left( \frac{\Psi_{2,-2}}{\lambda_{2,-2}} \right)\nonumber\\
&+&   \bar{d}_{1}^{2,3} \left( \frac{\Psi_{2,-3}}{\lambda_{2,-3}} \right) 
+   \bar{d}_{1}^{3,1} \left( \frac{\Psi_{3,-1}}{\lambda_{3,-1}} \right) + \bar{d}_{1}^{3,2} \left( \frac{\Psi_{3,-2}}{\lambda_{3,-2}} \right) + \bar{d}_{1}^{4,1} \left( \frac{\Psi_{4,-1}}{\lambda_{4,-1}} \right) \nonumber\\
&+&     \bar{d}_{1}^{4,2} \left( \frac{\Psi_{4,-2}}{\lambda_{4,-2}} \right) + \bar{d}_{1}^{4,3} \left( \frac{\Psi_{4,-3}}{\lambda_{4,-3}} \right)
+   \bar{d}_{1}^{5,2} \left( \frac{\Psi_{5,-2}}{\lambda_{5,-2}} \right) + \bar{d}_{1}^{6,3} \left( \frac{\Psi_{6,-3}}{\lambda_{6,-3}} \right). \label{def:phi[1]:resonance}
\end{eqnarray} 
It should be emphasized that all eigenvalues $\lambda_{m,n}$ listed in the above expression are nonzero so that $\phi_1$ is finite.  More importantly, all coefficients in the above expression are independent of the time so that the corresponding wave profile does {\em not}  grow in time!  Note that, like the initial guess $\phi_0$ defined by (\ref{def:phi[0]:2}),  the common solution $\phi_1$ given by (\ref{def:phi[1]:resonance}) has three unknown coefficients $\bar{A}_1, \bar{B}_1$ and $\bar{C}_1$, which can be determined similarly by avoiding the ``secular''  terms in $\phi_2$.   So, the above approach has general meanings.  Therefore,  in a  similar way, we can get $\eta_m$ and $\phi_m$ successively, where $m=1,2,3$ and so on.

\begin{table}[htdp]
\caption{Roots of Eq. (\ref{eq:ABC}) }
\begin{center}
\begin{tabular}{|l|r|r|r|} \hline\hline
Series number &		&		&		\\ 
of roots  ($K$)             & $\bar{A}_0$ &  $\bar{B}_0$  &   $\bar{C}_0$  \\
 \hline
1 (Group-I) 	&	-0.0156112	&	0.0282054	&	-0.0084973	\\
2  	&	-0.0156112	&	-0.0282054	&	0.0084973	\\
3	&	0.0156112	&	0.0282054	&	-0.0084973	\\
4	&	0.0156112	&	-0.0282054	&	0.0084973	\\ \hline\hline
5 (Group-II) 	&	-0.0155774	&	-0.0141927	&	-0.0113800	\\
6 	&	-0.0155774	&	 0.0141927	&	 0.0113800	\\
7	&	 0.0155774	&	-0.0141927	&	-0.0113800	\\
8	&	 0.0155774	&	0.0141927	&	 0.0113800	\\ \hline\hline
9 	(Group-III)	&	-0.0155626	&	0.0106109	&	-0.0226353	\\
10 &	-0.0155626	&	-0.0106109	&	0.0226353	\\
11	&	 0.0155626	&	0.0106109	&	-0.0226353	\\
12	&	 0.0155626	&	-0.0106109	&	 0.0226353	\\ \hline\hline
\end{tabular}
\end{center}
\label{Table:roots}
\end{table}

\begin{figure}
\setcaptionwidth{5in} \centering
\includegraphics[scale=0.4]{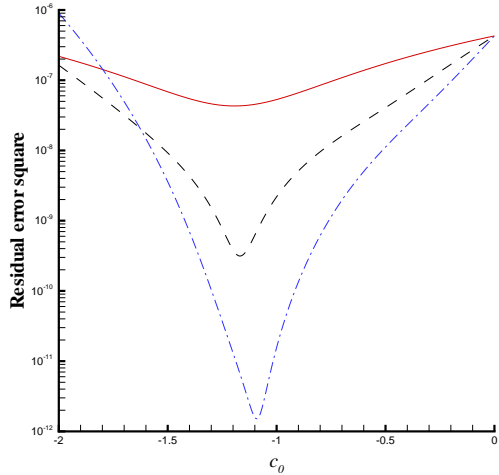}
\caption{Residual error square versus the convergence control parameter $c_0$ in case of (\ref{def:case:A}) with $k_2/k_1=0.8925$.}\label{figure:error:DR:4}
\end{figure}

\begin{figure}
\setcaptionwidth{5in} \centering
\includegraphics[scale=0.4]{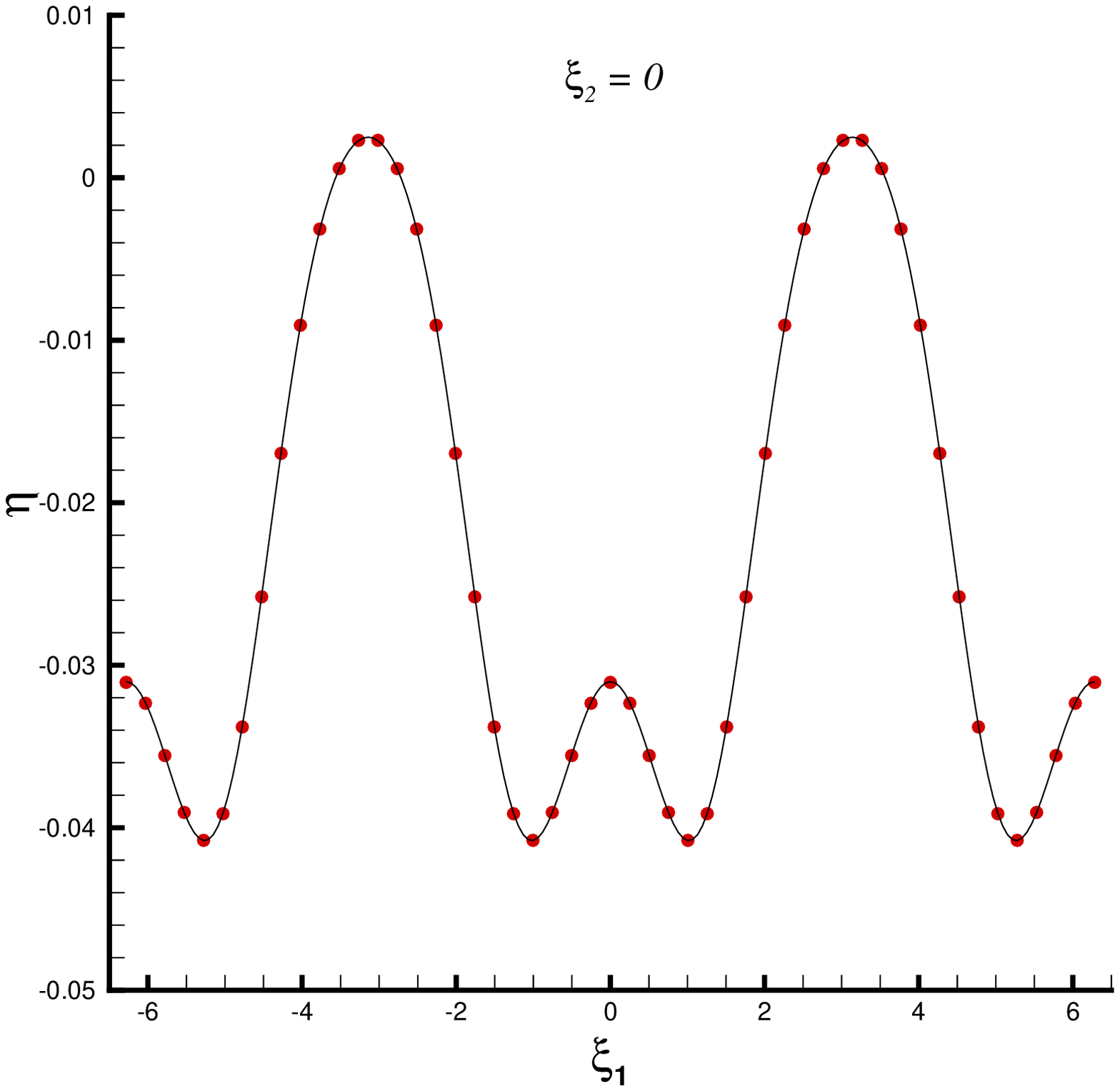}
\caption{Resonant wave profile at $\xi_2=0$  in case of (\ref{def:case:A}) with $k_2/k_1=0.8925$. Symbols: 5th-order approximation; Solid line: 10th-order approximation.}\label{figure:Eta:DR4a}

\setcaptionwidth{5in} \centering
\includegraphics[scale=0.4]{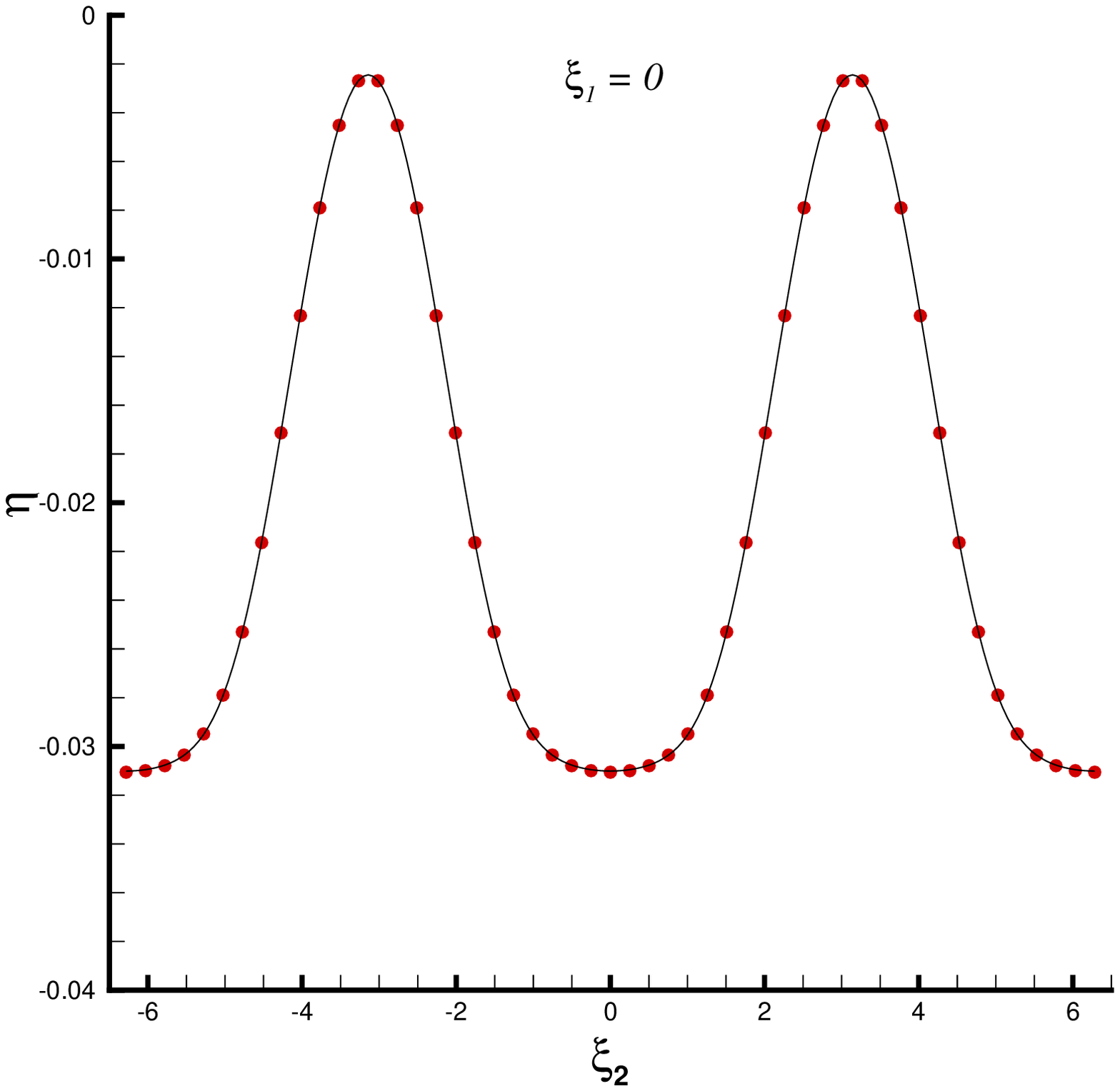}
\caption{Resonant wave profile at $\xi_1=0$  in case of (\ref{def:case:A}) with $k_2/k_1=0.8925$. Symbols: 5th-order approximation; Solid line: 10th-order approximation.}\label{figure:Eta:DR4b}
\end{figure}

Note that the wave amplitude components  $a_{1,0}$ and $a_{0,1}$ in Table~\ref{Table:wave-amplitude:all} are negative.  To calculate the corresponding wave amplitude  components  in case of (\ref{def:case:A})  with $k_2/k_1=0.8925$,  we choose the 2nd root in Group-I, i.e. 
\[   \bar{A}_0 = 	-0.0156112, \;\; \bar{B}_0 =	-0.0282054,	\;\; \bar{C}_0 = 	0.00849726.	 \]
Similarly, we can choose an optimal value of the so-called convergence-control parameter $c_0$ by plotting the curves of the residual error square ${\cal E}_m^\phi$ versus $c_0$, as shown in Fig.~\ref{figure:error:DR:4}, which indicates that  the series solution converges in the region $-1.6 < c_0 <0$ and that the optimal value of $c_0$   is close to -1.  For simplicity, we take $c_0=-1$.   The  residual error squares ${\cal E}_m^\phi$ and ${\cal E}_m^\eta$ of the two boundary conditions  decrease  rapidly to the level $10^{-19}$ (at the 20th-order approximation), as shown in Table~\ref{Table:error:resonance}.   According to the Convergence Theorem proved in Appendix B,  the homotopy-series (\ref{series:phi}) and (\ref{series:eta}) satisfy the original governing equation (\ref{geq:phi}) and all boundary conditions (\ref{bc:eta}), (\ref{bc:phi}) and (\ref{bc:phi:bottom}).   Besides,   it is found that  the corresponding wave amplitude components $a_{1,0}, a_{0,1}$ and $a_{2,-1}$ converge to -0.02051, -0.023212, 0.0089752, respectively, as shown in Table~\ref{Table:N=3:wave-amplitude}.   To confirm the convergence, we  further  employ the homotopy-Pad\'{e} technique \cite{Liao2007-SAM, LiaoBook2003}  to accelerate the convergence and obtain the same convergent wave amplitude  components 
\[ a_{1,0} = -0.0205119, \;\; a_{0,1} = -0.0232118, \;\; a_{2,-1} = 0.0089752,  \]
as shown in Table~\ref{Table:N=3:wave-amplitude:HP}.   Furthermore, the corresponding  wave profile converges quickly,  too, as shown in Figs.~\ref{figure:Eta:DR4a} and \ref{figure:Eta:DR4b}.   Therefore,  we indeed get convergent series solution of resonant waves with {\em constant} amplitudes  even when the resonant condition (\ref{resonance}) is exactly satisfied.    

\begin{figure}
\setcaptionwidth{5in} \centering
\includegraphics[scale=0.4]{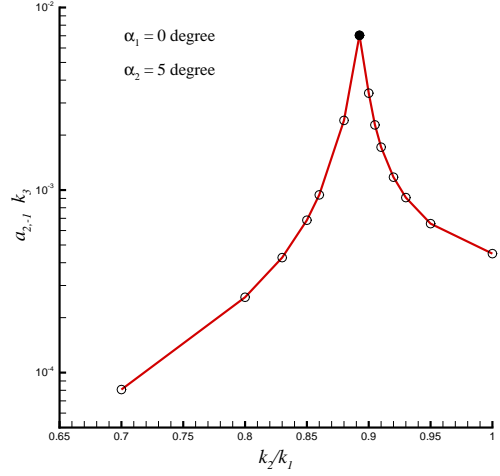}
\caption{The dimensionless wave amplitude  component $k_3 (a_{2,-1})$ versus $k_2/k_1$  in case of (\ref{def:case:A}).  Filled circle: result when $k_2/k_1 =0.8925$; Open circles:  results when $k_2/k_1 \neq 0.8925$. }\label{figure:a21-k}
\end{figure}

Combining the above result with those listed in Table~\ref{Table:wave-amplitude:all}, we obtain the whole pattern of the dimensionless wave amplitude  component $k_3  (a_{2,-1})$ versus $k_2/k_1$ in case of (\ref{def:case:A}), as shown in Fig.~\ref{figure:a21-k}.  It is true that, as the ratio $k_2/k_1$ goes to 0.8925, corresponding to the criterion (\ref{resonance}) of wave resonance, the dimensionless wave amplitude component  $k_3  \; a_{2,-1} $ arrives its maximum.   Besides,  the resonant wave profile in case of $k_2/k_1=0.8925$ becomes more complicated, if compared with the non-resonant one in case of $k_2/k_1=1$.    However, it should be emphasized that the amplitude $|a_{2,-1}|$ of  the wave  component $\cos(2\xi_1-\xi_2)$ is a {\em finite} constant,  even if the resonance condition  (\ref{resonance}) is satisfied exactly.    Besides,  it is surprising that, in case of  $k_2/k_1=0.8925$,   the amplitude $|a_{2,-1}|$ of the resonant wave component  is even much {\em smaller} than the wave amplitudes $|a_{1,0}|$ and $|a_{0,1}|$ of the two primary waves!

\begin{figure}
\setcaptionwidth{5in} \centering
\includegraphics[scale=0.4]{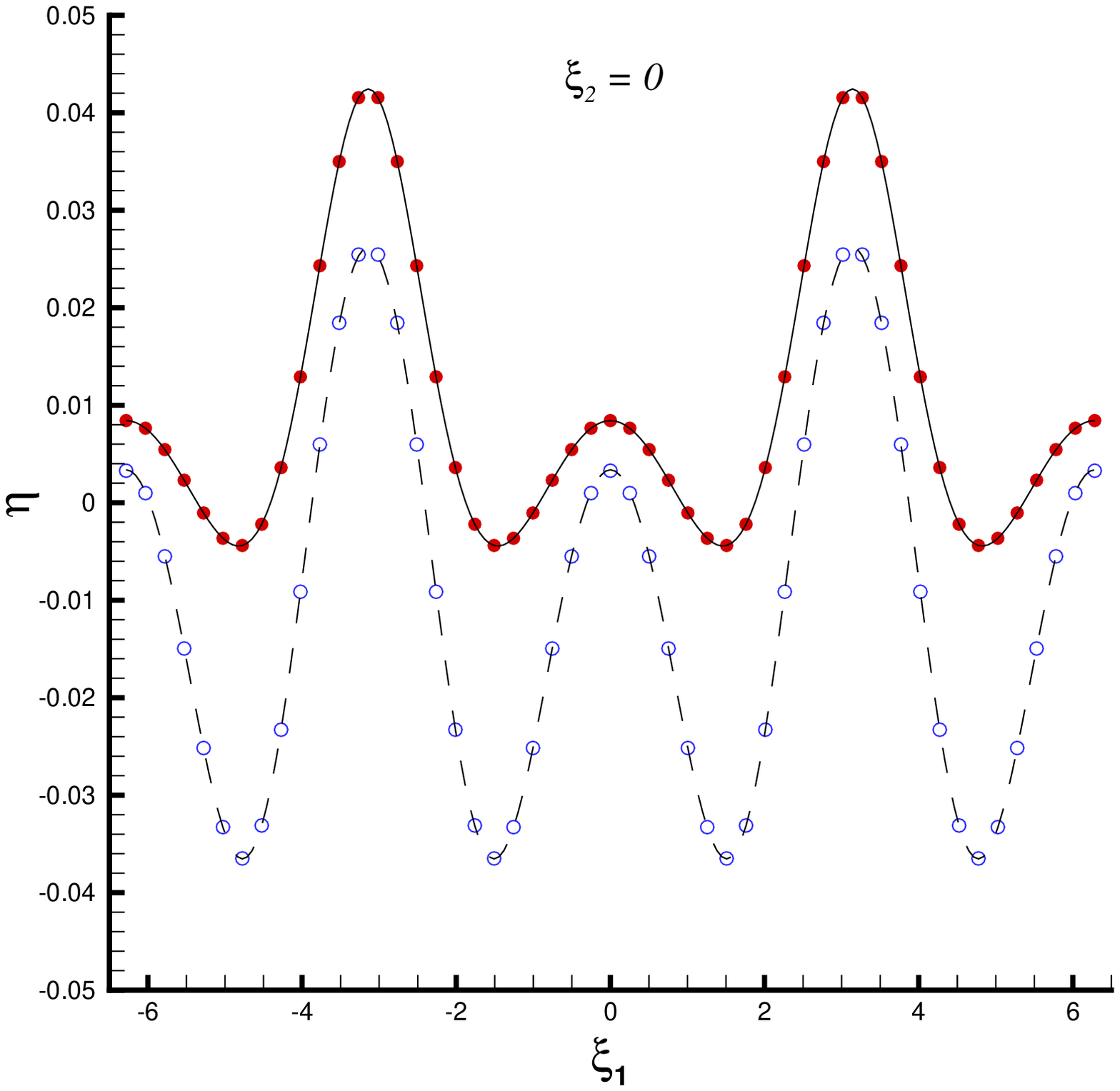}
\caption{Resonant wave profile at $\xi_2= 0$ in case of (\ref{def:case:A}) with $k_2/k_1=0.8925$.  Solid line: the  10th-order of approximation of Group II; Dashed line:  the 10th-order of approximation of Group III; Symbols: 6th-order approximations. }\label{figure:Eta:DR:68a}
\end{figure}

\begin{figure}
\setcaptionwidth{5in} \centering
\includegraphics[scale=0.4]{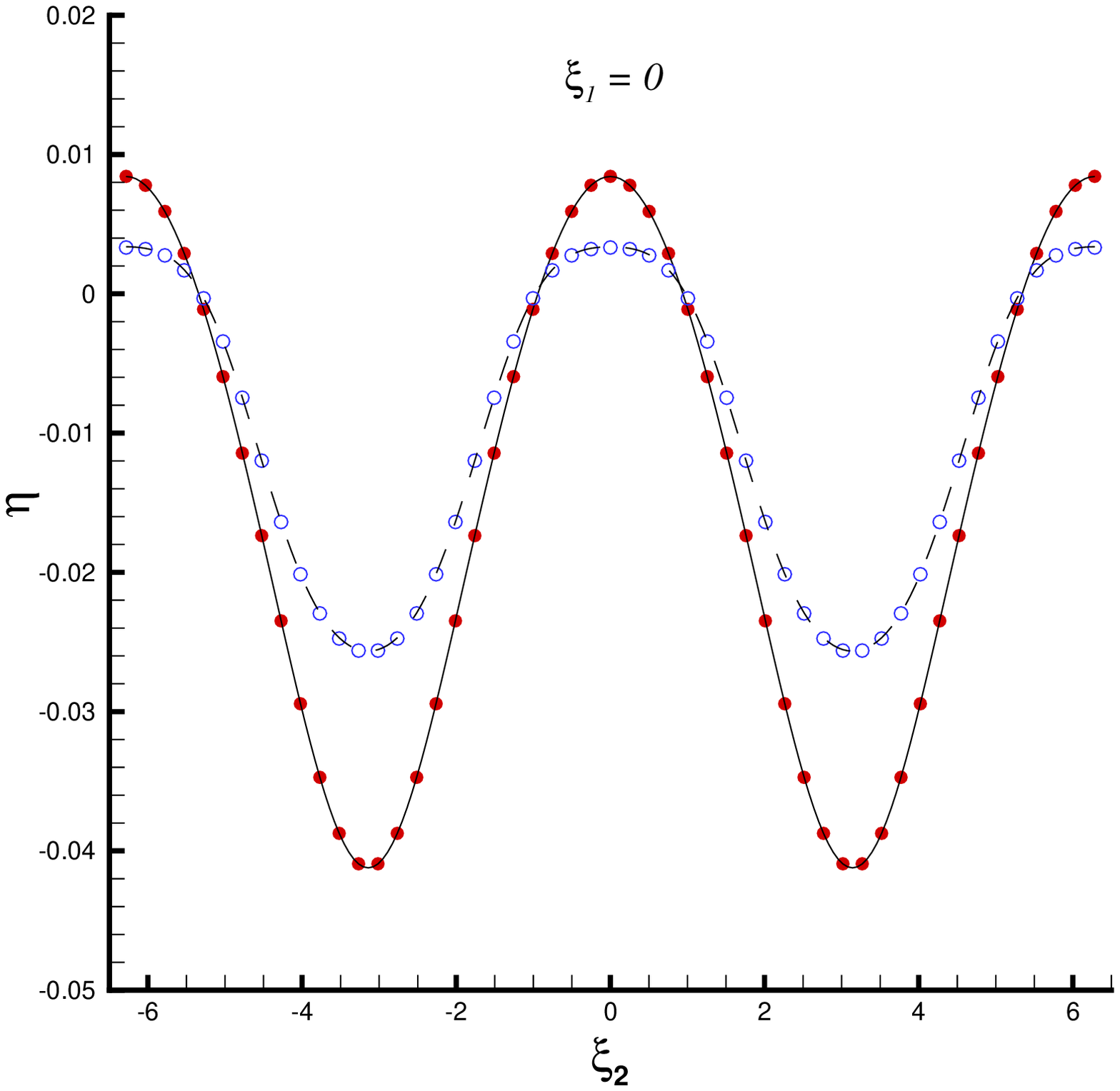}
\caption{Resonant wave profile at $\xi_1 = 0$ in case of (\ref{def:case:A}) with $k_2/k_1=0.8925$.   Solid line: the  10th-order of approximation of Group II; Dashed line:  the 10th-order of approximation of Group III; Symbols: 6th-order approximations . }\label{figure:Eta:DR:68b}
\end{figure}

The above results are obtained by using the 2nd root of Group I in Table~\ref{Table:roots}.  Similarly, using different roots in Table~\ref{Table:roots}, we can search for the corresponding convergent series solutions.  It is found that the four different roots of each group in Table~\ref{Table:roots} give the different wave amplitude components $a_{1,0}, a_{0,1}$ and $a_{2,-1}$.  But, they have the same absolute values $|a_{1,0}|, |a_{0,1}|$ and $|a_{2,-1}|$, as listed in Table~\ref{Table:wave-amplitude:group}.   Considering the fact that the wave  energy spectrum  is determined by the amplitude square of wave components,  we regard the four different solutions in each group as the same.   Thus, in case of (\ref{def:case:A}) with $k_2/k_1 = 0.8925$,  there are {\em three}  different  resonant-wave patterns with {\em different} wave energy spectrums.    The resonant wave profiles of Group II and III are as shown in Figs.~\ref{figure:Eta:DR:68a} and \ref{figure:Eta:DR:68b}.    Here, we would like to emphasize that  the amplitude $|a_{2,-1}|$ of the resonant wave is the smallest in Group I, and is the middle in Group II, although it is the largest in Group III.   So, the amplitude $|a_{2,-1}|$ of the resonant wave is not special at all: it is just normal as the wave amplitude components $|a_{1,0}|$ and $|a_{0,1}|$  of the two primary waves.   What we would like to emphasize here is that, for a fully developed wave system,  there exist {\em multiple} solutions  when the resonance condition is exactly satisfied.      Besides, the resonant wave amplitude may be much smaller than primary wave amplitudes.    These interesting results have not been reported, to the best of our knowledge. 

Let $\Pi$ denote the sum of amplitude square of all wave components  and  write
\[     \Pi_0 = a_{1,0}^2 +a_{0,1}^2+a_{2,-1}^2.\]
It is found that $\Pi_0/\Pi=98.82\%, 98.27\%$ and $99.76\%$ for Group I, II and III in case of (\ref{def:case:A}) with $k_2/k_1 = 0.8925$ for resonant waves, respectively.  This is mainly because amplitudes of other wave components such as $a_{1,-2}$ are much smaller, as shown in Table~\ref{Table:wave-amplitude:group}.  Thus, these three wave components nearly contain the whole wave energy.   Note that, given two primary traveling waves with wave numbers ${\bf k}_1$ and ${\bf k}_2$,   there exist an infinite number of different wave components $a_{m,n}\cos(m\xi_2+n\xi_2)$ with the wave number $ m {\bf k}_1 + n{\bf k}_2$,   where $m$ and $n$ are arbitrary integers.  Let
\[      {\bf K} = \left\{   m {\bf k}_1 + n{\bf k}_2 |  \mbox{$m,n$ are integers} \right\}      \]
denote a set of all these wave numbers.   Each wave number  in ${\bf K} $ corresponds to an eigenfunction defined by (\ref{def:psi}) with an eigenvalue defined by (\ref{def:eigenvalue}).     Our computations suggest that, for a fully developed wave system with {\em small} amplitudes,  the main of wave energy focuses on the wave components whose eigenvalues are zero (or close to zero).   This provides us an alternative  explanation for  the so-called  wave resonance.     According to this explanation,   the resonant wave $a_{2,-1} \cos(2\xi_1-\xi_2)$ is as important as the two primary waves $a_{1,0} \cos\xi_1$ and $ a_{0,1}\cos\xi_2$.   

\begin{figure}
\setcaptionwidth{5in} \centering
\includegraphics[scale=0.4]{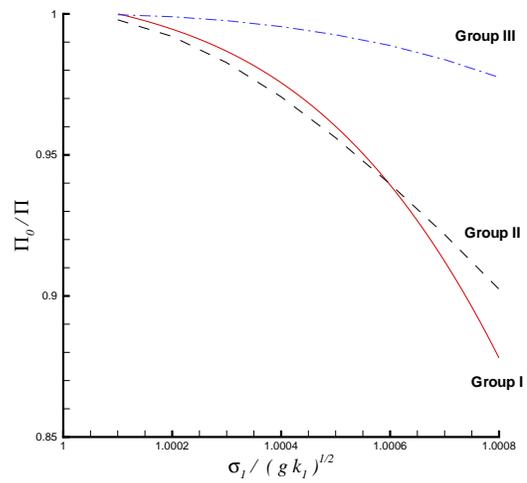}
\caption{$\Pi_0/\Pi$ versus $\sigma_1/\sqrt{g k_1}$ in case of (\ref{def:case:A}) when $k_2/k_1 =0.8925$ and $\sigma_1/\sqrt{g k_1}= \sigma_2/\sqrt{g k_2} $ for different group of solutions.  Solid line: Group I; Dashed line: Group II; Dash-dotted line: Group III. }\label{figure:energy:Pi[0]}
\end{figure}

The amplitudes of the resonant waves (related to Group I) in case of $\alpha_1=0,\alpha_2=\pi/36, k_2/k_1 = 0.8925$ with different ratios of  $\sigma_1/\sqrt{g k_1} = \sigma_2/\sqrt{g k_2}$ are as shown in Table~\ref{Table:amplitude:Group-1}.   The corresponding wave energy distributions are given in Table~\ref{Table:wave-energy:Group-1}.   It is found that the two primary waves contain most of the wave energy in this special case.   Besides, as the ratio of $\sigma_1/\sqrt{g k_1} =\sigma_2/\sqrt{g k_2}$ increases, the resonant wave $a_{2,-1} \cos(2\xi_1-\xi_2)$  contains less and less percentage of the whole wave energy.   Especially,  when $\sigma_1/\sqrt{g k_1} =\sigma_2/\sqrt{g k_2} = 1.0008$, the resonant wave of Group I contains only 2.22\% of the whole wave energy.   This result is interesting, but a little surprising, because the resonance wave is traditionally supposed to have a large wave amplitude and thus to contain the main of wave energy.  We will attempt to explain this phenomena in \S 4.   The amplitudes of the resonant waves related to Group II are listed in Table~\ref{Table:amplitude:Group-2} and the corresponding wave energy distribution is given in Table~\ref{Table:wave-energy:Group-2}.   It is found that the resonant waves of  Group II have the comparable wave amplitudes with the comparable  percentage of wave energy to one of primary waves.    
 Among three groups, there exists only one group (i.e. Group III) such that  the resonant waves have the largest wave amplitude and besides contain the main part of wave energy,   as shown in Table~\ref{Table:amplitude:Group-3}  and Fig.~\ref{figure:energy:a[2,1]}.    Note that,  the primary and resonant waves contain the most part of wave energy, especially when all wave amplitude components are very small, as shown in Fig.~\ref{figure:energy:Pi[0]}.   However, as the wave amplitudes increase,  the primary and resonant waves contain less and less   percentage of wave energy, as shown in Fig.~\ref{figure:energy:Pi[0]}.   This indicates that Phillips' wave resonance condition (\ref{resonance:Phillips}) might hold only for small-amplitude traveling waves.

The above results have general meaning, although they are obtained in a special case (\ref{def:case:A}).   
These results  strongly suggest  that,  for a fully developed system of two traveling waves,  all wave amplitudes do  {\em not} grow linearly in time even if the wave resonance condition is exactly satisfied.    This conclusion is also true for arbitrary number of traveling waves, as shown below.   Currently, by means of DNS (direct numerical simulation) of the evolution of nonlinear random water waves fields with a continuous spectrum,  Annenkov {\em et al}  \cite{Annenkov2006JFM} investigated the role of exactly resonant, nearly resonant and non-resonant wave interactions, and their results indicate that the amplitudes of wave packets tend to constants.   Their results, although obtained for a continuous wave spectrum,  support our conclusions mentioned above.  

\begin{table}[htdp]
\caption{Residual error square of the two boundary conditions in case of (\ref{def:case:A}) with $k_2/k_1=0.8925$ }
\begin{center}
\begin{tabular}{|c|c|c|}\hline\hline
$m$	&	${\cal E}_m^\phi$	&	${\cal E}_m^\eta$	\\ \hline
1	&	4.0 $\times 10^{-7}$	&	4.8$\times 10^{-4}$	\\
3	&	1.8$\times 10^{-8}$	&	1.0$\times 10^{-6}$	\\
5	&	2.4 $\times 10^{-10}$	&	1.2$\times 10^{-8}$	\\
8	&	1.2 $\times 10^{-13}$	&	1.5$\times 10^{-11}$	\\
10	&	1.3 $\times 10^{-14}$	&	1.2$\times 10^{-12}$	\\
15	&	1.2 $\times 10^{-17}$	&	3.4$\times 10^{-16}$	\\
20	&	1.2 $\times 10^{-20}$	&	1.9$\times 10^{-19}$	\\ \hline\hline
\end{tabular}
\end{center}
\label{Table:error:resonance}
\end{table}

\begin{table}[htdp]
\caption{Analytic approximations of wave amplitude components in case of (\ref{def:case:A}) with $k_2/k_1=0.8925$ }
\begin{center}
\begin{tabular}{|c|c|c|c|} \hline\hline
Order of appr.  &	$a_{1,0}$	&	 $a_{0,1}$ 	&	$a_{2,-1}$   \\  \hline
1	&	-0.015616	&	-0.028214	&	0.0084999	\\
3	&	-0.020587	&	-0.023469	&	0.0094143	\\
5	&	-0.020533	&	-0.023231	&	0.0090083	\\
7	&	-0.020504	&	-0.023220	&	0.0089755	\\
9	&	-0.020511	&	-0.023213	&	0.0089757	\\
11	&	\underline{-0.020512}	&	\underline{-0.023212}	&	0.0089753	\\
13	&	-0.020512	&	-0.023212	&	\underline{0.0089752}	\\
15	&	-0.020512	&	-0.023212	&	0.0089752	\\
18	&	-0.020512	&	-0.023212	&	0.0089752	\\
\hline\hline
\end{tabular}
\end{center}
\label{Table:N=3:wave-amplitude}
\end{table}

\begin{table}[htdp]
\caption{Analytical approximations of wave amplitude components given by the $[m,m]$ homotopy-Pad\'{e}  method in case of   (\ref{def:case:A}) with $k_2/k_1=0.8925$}
\label{Table:N=3:wave-amplitude:HP}
\begin{center}
\begin{tabular}{|c|c|c|c|}\hline\hline 
$m$	&	$a_{1,0}$	& $a_{0,1}$	&	$a_{2,-1}$ \\  \hline
2	&	-0.0206010	&	-0.0232058	&	0.0089660	\\
3	&	-0.0204911	&	-0.0232251	&	0.0089568	\\
4	&	-0.0205102	&	-0.0232172	&	0.0089780	\\
5	&	-0.0205122	&	\underline{-0.0232118}	&	\underline{0.0089752}	\\
6	&	\underline{-0.0205119}	&	-0.0232118	&	0.0089752	\\
7	&	-0.0205119	&	-0.0232118	&	0.0089752	\\
8	&	-0.0205119	&	-0.0232118	&	0.0089752	\\
9	&	-0.0205119	&	-0.0232118	&	0.0089752	\\
\hline\hline
\end{tabular}
\end{center}

\caption{Multiple amplitudes of resonant waves in case of (\ref{def:case:A}) with $k_2/k_1=0.8925$ given by different roots of Eq. (\ref{eq:ABC}) listed in Table~\ref{Table:roots}. }
\label{Table:wave-amplitude:group}
\begin{center} 
\begin{tabular}{|c|c|c|c|c|} \hline\hline
	&	$|a_{1,0}|$	&	$|a_{0,1}|$	&	$|a_{2,-1}|$	&	$|a_{1,-2}|$ 	\\ \hline
Group I	&	0.02051186921	&	0.02321179687	&	0.00897520547	&	0.00022907754	\\
Group II	&	0.01475607438	&	0.01002488089	&	0.01464333477	&	0.00096077706	\\
Group III	&	0.00971236473	&	0.01032248128	&	0.02576462018	&	0.00059968416	\\ 
\hline\hline
\end{tabular}
\end{center}
\end{table}

\begin{table}[htdp]
\caption{Amplitudes of resonant waves ( Group I )  in case of $\alpha_1=0, \alpha_2=\pi/36, k_2/k_1=0.8925$ with different ratios of $\sigma_1/\sqrt{g k_1} = \sigma_2/\sqrt{g k_2}$.}
\begin{center}
\begin{tabular}{|c|c|c|c|} \hline\hline
$\sigma_1/\sqrt{g k_1}, \sigma_2/\sqrt{g k_2}$  	&	$|a_{1,0}|$	&	$|a_{0,1}|$	&	$|a_{2,-1}|$	\\  \hline
1.0001	&	0.0117	&	0.0141	&	0.0058	\\
1.0002	&	0.0166	&	0.0195	&	0.0078	\\
1.0003	&	0.0205	&	0.0232	&	0.0090	\\
1.0004	&	0.0238	&	0.0259	&	0.0095	\\
1.0005	&	0.0266	&	0.0278	&	0.0096	\\
1.0006	&	0.0290	&	0.0291	&	0.0092	\\
1.0007	&	0.0312	&	0.0301	&	0.0085	\\
1.0008	&	0.0331	&	0.0308	&	0.0076	\\
\hline\hline
\end{tabular}
\end{center}
\label{Table:amplitude:Group-1}
\end{table}

\begin{table}[htdp]
\caption{Amplitudes of resonant waves ( Group II )  in case of $\alpha_1=0, \alpha_2=\pi/36, k_2/k_1=0.8925$ with different ratios of $\sigma_1/\sqrt{g k_1} = \sigma_2/\sqrt{g k_2}$.}
\begin{center}
\begin{tabular}{|c|c|c|c|} \hline\hline
$\sigma_1/\sqrt{g k_1}, \sigma_2/\sqrt{g k_2}$  	&	$|a_{1,0}|$	&	$|a_{0,1}|$	&	$|a_{2,-1}|$	\\  \hline
1.0001	&	0.0090	&	0.0063		&	0.0084	\\
1.0002	&	0.0124	&	0.0085		&	0.0119	\\
1.0003	&	0.0148	&	0.0100		&	0.0146	\\
1.0004	&	0.0166	&	0.0111		&	0.0169	\\
1.0005	&	0.0181	&	0.0120		&	0.0188	\\
1.0006	&	0.0194	&	0.0126		&	0.0206	\\
1.0007	&	0.0204	&	0.0130		&	0.0222	\\
1.0008	&	0.0214 	&	0.0134		&	0.0237	\\
\hline\hline
\end{tabular}
\end{center}
\label{Table:amplitude:Group-2}
\end{table}

\begin{table}[htdp]
\caption{Amplitudes of resonant waves ( Group III )  in case of $\alpha_1=0, \alpha_2=\pi/36, k_2/k_1=0.8925$ with different ratios of $\sigma_1/\sqrt{g k_1} = \sigma_2/\sqrt{g k_2}$.}
\begin{center}
\begin{tabular}{|c|c|c|c|} \hline\hline
$\sigma_1/\sqrt{g k_1}, \sigma_2/\sqrt{g k_2}$  	&	$|a_{1,0}|$	&	$|a_{0,1}|$	&	$|a_{2,-1}|$	\\  \hline
1.0001	&	0.0060	&	0.0060		&	0.0148	\\ 
1.0002	&	0.0082	&	0.0085		&	0.0210	\\
1.0003	&	0.0097	&	0.0103		&	0.0258	\\
1.0004	&	0.0108	&	0.0118		&	0.0298	\\
1.0005	&	0.0116	&	0.0131		&	0.0333	\\
1.0006	&	0.0122	&	0.0143		&	0.0365	\\
1.0007	&	0.0125	&	0.0153		&	0.0394	\\
1.0008	&	0.0127 	&	0.0163 		&	0.0420  \\
\hline\hline
\end{tabular}
\end{center}
\label{Table:amplitude:Group-3}
\end{table}

\begin{table}[htdp]
\caption{Wave energy distribution of resonant waves ( Group I )  in case of $\alpha_1=0, \alpha_2=\pi/36, k_2/k_1=0.8925$ with different ratios of $\sigma_1/\sqrt{g k_1} = \sigma_2/\sqrt{g k_2}$.}
\begin{center}
\begin{tabular}{|c|l|l|l|l|} \hline\hline
$\sigma_1/\sqrt{g k_1}, \sigma_2/\sqrt{g k_2}$  	&	$a^2_{1,0}/\Pi$	&	$a^2_{0,1}/\Pi$	&	$a^2_{2,-1}/\Pi$	&	$\Pi_0/\Pi$	\\  \hline
1.0001	&	36.97\%	&	 53.95\% &	8.99\%	&	99.91\%	\\
1.0002	&	38.40\%	&	52.74\%	&	8.42\%	&	99.56\%	\\
1.0003	&	39.97\%	&	51.19\%	&	7.65\%	&	98.82\%	\\	
1.0004	&	41.59\%	&	49.28\%	&	6.70\%	&	97.57\%	\\
1.0005	&	43.10\%	&	47.09\%	&	5.60\%	&	95.78\%	\\
1.0006	&	44.38\%	&	44.75\%	&	4.43\%	&	93.56\%	\\
1.0007	&	45.65\%	&	42.73\%	&	3.58\%	&	91.95\%	\\
1.0008	&	46.23\%	&	39.07\%	&	2.22\%	&	87.52\%	\\		
\hline\hline
\end{tabular}
\end{center}
\label{Table:wave-energy:Group-1}
\end{table}

\begin{table}[htdp]
\caption{Wave energy distribution of resonant waves ( Group II )  in case of $\alpha_1=0, \alpha_2=\pi/36, k_2/k_1=0.8925$ with different ratios of $\sigma_1/\sqrt{g k_1} = \sigma_2/\sqrt{g k_2}$.}
\begin{center}
\begin{tabular}{|c|l|l|l|l|} \hline\hline
$\sigma_1/\sqrt{g k_1}, \sigma_2/\sqrt{g k_2}$  	&	$a^2_{1,0}/\Pi$	&	$a^2_{0,1}/\Pi$	&	$a^2_{2,-1}/\Pi$	&	$\Pi_0/\Pi$	\\  \hline
1.0001	&	42.30\%	&	 20.43\% &	37.07\%	&	99.79\%	\\
1.0002	&	41.26\%	&	19.50\%	&	38.44\%	&	99.20\%	\\
1.0003	&	40.17\%	&	18.54\%	&	39.56\%	&	98.27\%	\\	
1.0004	&	39.05\%	&	17.56\%	&	40.44\%	&	97.06\%	\\
1.0005	&	37.93\%	&	16.56\%	&	41.12\%	&	95.61\%	\\
1.0006	&	36.79\%	&	15.54\%	&	41.64\%	&	93.97\%	\\
1.0007	&	35.65\%	&	14.51\%	&	42.02\%	&	92.18\%	\\
1.0008	&	34.49\%	&	13.46\%	&	42.29\%	&	90.24\%	\\		
\hline\hline
\end{tabular}
\end{center}
\label{Table:wave-energy:Group-2}
\end{table}

\begin{table}[htdp]
\caption{Wave energy distribution of resonant waves ( Group III )  in case of $\alpha_1=0, \alpha_2=\pi/36, k_2/k_1=0.8925$ with different ratios of $\sigma_1/\sqrt{g k_1} = \sigma_2/\sqrt{g k_2}$.}
\begin{center}
\begin{tabular}{|c|l|l|l|l|} \hline\hline
$\sigma_1/\sqrt{g k_1}, \sigma_2/\sqrt{g k_2}$  	&	$a^2_{1,0}/\Pi$	&	$a^2_{0,1}/\Pi$	&	$a^2_{2,-1}/\Pi$	&	$\Pi_0/\Pi$	\\  \hline
1.0001	&	12.13\%	&	12.47\% &	75.37\%	&	99.97\%	\\
1.0002	&	11.53\%	&	12.38\%	&	75.99\%	&	99.90\%	\\
1.0003	&	10.88\%	&	12.29\%	&	76.58\%	&	99.76\%	\\	
1.0004	&	10.20\%	&	12.20\%	&	77.15\%	&	99.55\%	\\
1.0005	&	 9.47\%	&	12.11\%	&	77.68\%	&	99.26\%	\\
1.0006	&	 8.72\%	&	12.01\%	&	78.15\%	&	98.88\%	\\
1.0007	&	 7.94\%	&	11.90\%	&	78.53\%	&	98.38\%	\\
1.0008	&	 7.15\%	&	11.78\%	&	78.82\%	&	97.75\%	\\		
\hline\hline
\end{tabular}
\end{center}
\label{Table:wave-energy:Group-3}
\end{table}

\begin{figure}
\setcaptionwidth{5in} \centering
\includegraphics[scale=0.4]{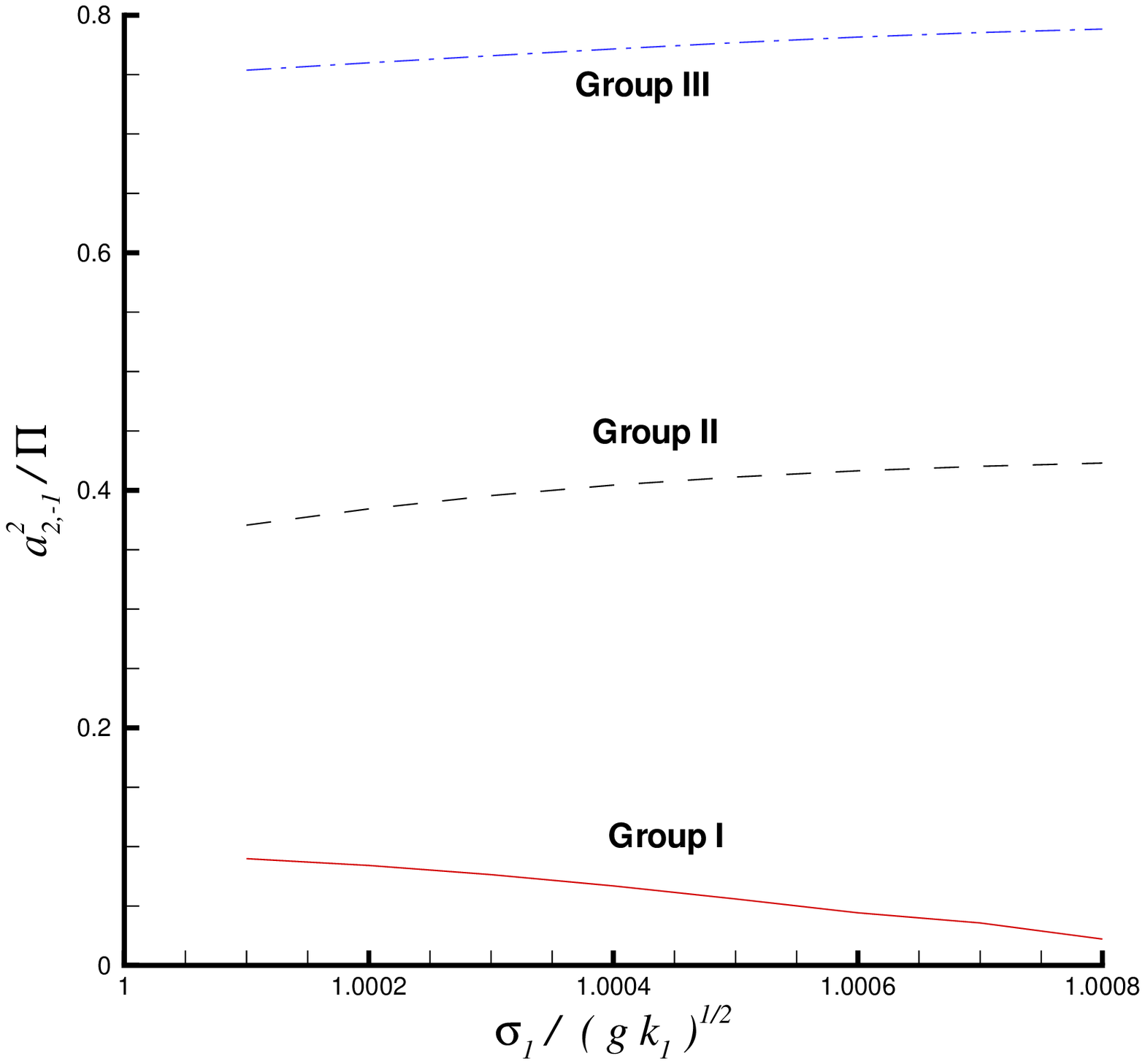}
\caption{$a^2_{2,-1}/\Pi$ versus $\sigma_1/\sqrt{g k_1}$ in case of (\ref{def:case:A}) when $k_2/k_1 =0.8925$ and $\sigma_1/\sqrt{g k_1}= \sigma_2/\sqrt{g k_2} $ for different group of solutions.  Solid line: Group I; Dashed line: Group II; Dash-dotted line: Group III. }\label{figure:energy:a[2,1]}
\end{figure}

\clearpage\newpage

\section{Resonance condition of arbitrary number of waves}

\subsection{Resonance condition for small-amplitude waves}

In \S 3, we shaw that, for a fully developed system of two primary traveling waves,  Phillips' resonance condition for small-amplitude waves is exactly  equivalent to the zero eigenvalue of the eigenfunction related to the resonant wave.   This conclusion has general meanings, and can be  easily expanded to give a resonance condition for {\em arbitrary} number of travel waves with small amplitude.   The key is to give an explicit expression of the eigenvalue in case of arbitrary number of traveling waves.       

Let us consider the nonlinear interaction of $\kappa$ periodic traveling waves with small amplitudes, where $2 \leq \kappa < +\infty$  is an arbitrary integer.  Let ${\bf k}_n$ and $\sigma_n$  ($1\leq n \leq \kappa$) denote the given wave number and angular frequency of the $n$th periodic traveling waves in deep water.     Define the variable
\[
\xi_n = {\bf k}_m \cdot {\bf r} -\sigma_n \; t,
\]
which has clear physical meanings.   Then, 
\[    \varphi(x,y,z,t) = \phi(\xi_1,\xi_2,\cdots, \xi_\kappa,z), \;\; \zeta(x,y,t) = \eta(\xi_1,\xi_2,\cdots, \xi_\kappa).   \]
Similarly,  we have
\begin{eqnarray}
\frac{\partial^2\varphi}{\partial t^2} &=& \sum_{m=1}^{\kappa}\sum_{n=1}^{\kappa}\sigma_m \sigma_n  \; \frac{\partial^2 \phi}{\partial \xi_m \partial \xi_n}, \\
\nabla \varphi &=& \left( \sum_{m=1}^{\kappa} {\bf k}_m \; \frac{\partial \phi}{\partial \xi_m}\right) + {\bf k} \frac{\partial \phi}{\partial z} = {\bf u} = \hat\nabla \phi,\\
\nabla^2\varphi &=& \left( \sum_{m=1}^{\kappa}\sum_{n=1}^{\kappa}{\bf k}_m \cdot {\bf k}_n  \; \frac{\partial^2 \phi}{\partial \xi_m \partial \xi_n}\right)  + \frac{\partial^2 \phi}{\partial z^2} = \hat\nabla^2\phi,\\
\hat\nabla\phi \cdot \hat\nabla\phi &=& \left( \sum_{m=1}^{\kappa}\sum_{n=1}^{\kappa} {\bf k}_m \cdot {\bf k}_n \; \frac{\partial \phi}{\partial \xi_m} \frac{\partial \phi}{\partial \xi_n}\right)  + \left( \frac{\partial \phi}{\partial z}\right)^2 = {\bf u}^2 ,\\
\hat\nabla\phi \cdot \hat\nabla\psi &=& \left( \sum_{m=1}^{\kappa}\sum_{n=1}^{\kappa} {\bf k}_m \cdot {\bf k}_n \; \frac{\partial \phi}{\partial \xi_m} \frac{\partial \psi}{\partial \xi_n}\right)  +  \frac{\partial \phi}{\partial z} \frac{\partial \psi}{\partial z}.
\end{eqnarray}
Note that 
\begin{equation}
\Psi_{m_1,m_2,\cdots,m_\kappa, z} = \exp\left( \left|\sum_{n=1}^{\kappa} m_n {\bf k}_n\right|z\right)\; \sin \left( \sum_{n=1}^{\kappa} m_n \xi_n\right) \label{def:eigenfunction:general}
\end{equation}
satisfies the Laplace equation $\hat\nabla\phi=0$, i.e. 
\[    \hat\nabla^2 \Psi_{m_1,m_2,\cdots,m_\kappa} = 0. \]
 The two nonlinear boundary conditions on the free surface can be written by means of the operators defined above in a similar way.   Then, we can construct the zeroth-order deformation equations and the corresponding high-order deformation equations in a similar way as mentioned in \S 3.    Although it seems that the governing equations and boundary conditions become much more complicated in form than the original ones by means of these variables,  these multiple variables have very clear physical meanings which in fact greatly simplify solving the problem, as described below.     

Similarly,  we choose such an auxiliary linear operator
\begin{equation}
{\cal L}\phi = \left( \sum_{m=1}^{\kappa}\sum_{n=1}^{\kappa} \bar\sigma_m \; \bar\sigma_n \frac{\partial^2 \phi}{\partial \xi_m \partial \xi_n} \right)+ g \frac{\partial^2 \phi}{\partial z^2},  \label{def:L:general}
\end{equation}
where
\[   \bar\sigma_m = \sqrt{g\;k_m},  \;\;  k_m = |{\bf k}_m|\]
is based on the linear theory for small-amplitude waves.    The above auxiliary linear operator satisfies
\begin{equation}
{\cal L} \left( \Psi_{m_1,m_2,\cdots,m_\kappa}  \right) = \lambda_{m_1,m_2,\cdots,m_\kappa} \; \Psi_{m_1,m_2,\cdots,m_\kappa},\label{geq:eigenfunction:general}
\end{equation}
where
\begin{equation}
\lambda_{m_1,m_2,\cdots,m_\kappa} = g \left|\sum_{n=1}^{\kappa} m_n {\bf k}_n\right| - \left(  \sum_{n=1}^{\kappa} m_n \bar\sigma_n \right)^2 \label{def:eigenvalue:general}
\end{equation}
is the eigenvalue and $\Psi_{m_1,m_2,\cdots,m_\kappa} $  defined by (\ref{def:eigenfunction:general}) is the eigenfunction of the linear operator ${\cal L}$ defined by (\ref{def:L:general}).  Similarly, the inverse operator of (\ref{def:L:general}) satisfies 
\begin{equation}
{\cal L}^{-1} \left( \Psi_{m_1,m_2,\cdots,m_\kappa}  \right) = \frac{\Psi_{m_1,m_2,\cdots,m_\kappa}}{ \lambda_{m_1,m_2,\cdots,m_\kappa}}, \hspace{1.0cm} \lambda_{m_1,m_2,\cdots,m_\kappa} \neq 0.  \label{L:inverse:general}
\end{equation}

Note that the inverse operator (\ref {L:inverse:general}) has definition only for non-zero eigenvalue $\lambda_{m_1,m_2,\cdots,m_\kappa}$.   Besides, the eigenvalues of all  primary traveling waves are zero, i.e. there exist at least $\kappa$ zero eigenvalues for $\kappa$ primary waves.   Thus, the so-called wave resonance occurs when there are more than $\kappa$ zero eigenvalues.    So, enforcing $\lambda_{m_1,m_2,\cdots,m_\kappa}=0$ gives the resonance condition  
\begin{equation}
 g \left|\sum_{n=1}^{\kappa} m_n {\bf k}_n\right| = \left(  \sum_{n=1}^{\kappa} m_n \bar\sigma_n \right)^2,
 \hspace{1.0cm} \sum_{n=1}^{\kappa} m_n^2 >1,    \label{resonance:general}
\end{equation}
where $\bar\sigma_n = \sqrt{g \; k_n}$ with $k_n = |{\bf k}_n|$ is based on the linear theory for small-amplitude waves. 
Note that  (\ref{resonance}) is a special case of the above resonance condition.   Besides, the above formula contains the resonance condition given by Phillips \cite{Phillips1960JFM} and  thus is more general.  

  Assume that, for given $\kappa$ primary traveling waves, there are $N_\lambda \geq \kappa$ eigenfunctions whose eigenvalues are zero.   When $N_\lambda = \kappa$, there is no wave resonance.  However, when $N_\lambda > \kappa$, wave resonance occurs:  the wave energy transfers  greatly between the resonant wave and primary ones.    For simplicity, let $\Psi^*_m$ ($1\leq m \leq N_\lambda$) denote the $m$th eigenfunction with zero eigenvalue.  According to (\ref{geq:eigenfunction:general}),  it holds 
\begin{equation}
{\cal L} \left( \sum_{m=1}^{N_\lambda} A_m \; \Psi_m^* \right)  = 0 , \hspace{1.0cm}  N_\lambda \geq \kappa  \label{property:L:general}
\end{equation}
for any constant $A_m$.   So, we can always  choose such an initial guess that
\begin{equation}
\phi_0 = \sum_{m=1}^{N_\lambda} B_{0,m} \; \Psi_m^* , \label{def:phi[0]:general}
\end{equation}
where $B_{0,m}$ is unknown.   Similarly,  the $N_\lambda$ unknown constants $B_{0,m}$ ($1\leq m \leq N_\lambda$)  are determined by avoiding the ``secular'' terms in $\phi_1$.   Besides,  owing  to 
(\ref{property:L:general}), the common solution of $\phi_1$ contains $N_\lambda$ unknown constants $B_{1,m}$ ($1\leq m\leq N_\lambda$), which are similarly determined by avoiding the ``secular'' terms in $\phi_2$.  In this way,  one can solve the related high-order deformation equations successively, and  an optimal value of convergence-control parameter $c_0$ can be chosen so as to ensure the homotopy-series convergent quickly.     In theory,  the above approach is general, and works for arbitrary number of  primary periodic traveling  waves with small amplitudes.    It provides us a new way to investigate the weakly nonlinear interactions of more than four primary traveling waves with small amplitudes.

\subsection{Resonance condition for large wave-amplitude}

Note that  the general wave resonance condition (\ref{resonance:general}) holds only in case of $\bar\sigma_n = \sqrt{g \; k_n}$ with $k_n = |{\bf k}_n|$,  corresponding to small-amplitude gravity waves.   What is the resonance condition for arbitrary number of  traveling gravity waves with large-amplitude?  

\begin{figure}[h]
\setcaptionwidth{5in} \centering
\includegraphics[scale=0.4]{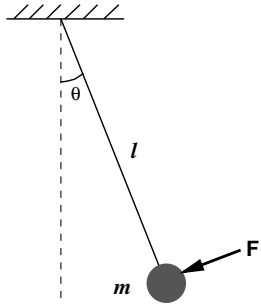}
\caption{The resonance of a simple pendulum }\label{figure:pendulum}
\end{figure}

To answer this question, we should consider the physical meanings of (\ref{resonance:general}).  In general,  the so-called resonance of a dynamic system occurs when the frequency of an external force (or disturbance) equals to the ``natural'' frequency of the dynamic system.    For example, let us consider the resonance of a simple pendulum, as shown in Fig.~\ref{figure:pendulum},  where ${\bf F} = A \; \cos(\omega t + \alpha)$ is the external force with the frequency $\omega$ and the phase difference $\alpha$.   When the maximum angle of oscillation $\theta_{max}$ is  so small that $\sin\theta \approx \theta$,  the simple pendulum has a natural frequency 
$\omega_0 \approx  \sqrt{g/l}$.  So,  if the frequency $\omega$ of the external force ${\bf F}$ is equal to the natural frequency $\omega_0$ of the simple pendulum,  i.e. $\omega = \sqrt{g/l}$, the total energy of the pendulum (and therefore $\theta_{max}$) quickly increases in case of the phase difference $\alpha = 0$ (or decreases in case of $\alpha = \pi$):  the so-called resonance occurs.     However,  $\omega_0 \approx  \sqrt{g/l}$ is only valid for small $\theta_{max}$:  the natural frequency $\omega_0$ increases as $\theta_{max}$ becomes larger.   So, as  the maximum angle of oscillation $\theta_{max}$  becomes so large that the natural frequency $\omega_0$  departs more and more  from the frequency $\omega = \sqrt{g/l}$  of the external force ${\bf F}$, then the simple  pendulum  gains less and less energy from the external  force:  the maximum angle of oscillation $\theta_{max}$  stops increasing when the simple pendulum can not gain energy from ${\bf F}$ any more in a period of oscillation.

The phenomenon of gravity wave resonance is physically similar to it in essence.   For a single traveling wave with the wave number ${\bf k}'$ and the ``natural'' angular frequency $\sigma'_0$,  the resonance occurs when there exists an ``external'' periodic disturbance  with the same  angular  frequency $\sigma'$, i.e. $\sigma' = \sigma'_0$.    It should be emphasized that  this resonance mechanism is physically reasonable even for large wave amplitude.

Let us consider  $\kappa$ primary traveling waves with wave number ${\bf k}_n$ and angular frequency $\sigma_n$, where $1\leq n \leq \kappa$.   Due to nonlinear interaction,  there exist a system of an infinite number of wave components 
\[     \cos\left(  \sum_{n=1}^{\kappa}  m_n \; \xi_n \right),    \]
where $m_n$ is an integer that can be negative, zero, or positive.    Note that 
\[  \sum_{n=1}^{\kappa}  m_n \; \xi_n   =  \left(  \sum_{n=1}^{\kappa}  m_n {\bf k}_{n} \right) \cdot {\bf r} 
- \left( \sum_{n=1}^{\kappa} m_n \; \sigma_n \right)  t. \]
So, 
\begin{equation}
  {\bf k}' = \sum_{n=1}^{\kappa} m_n \; {\bf k}_n   \label{def:k'}
\end{equation}  
is the wavenumber  and
\begin{equation}  
\sigma' = \left|   \sum_{n=1}^{\kappa} m_n \; \sigma_n  \right|     \label{def:sigma'}
\end{equation}
is the corresponding angular frequency of the nonlinear-interaction wave.  For the sake of simplicity, we call 
$   {\bf k}'   $
{\em the nonlinear-interaction wavenumber} and 
$    \sigma'   $  
{\em the nonlinear-interaction angular frequency}, respectively, where $\sum\limits_{n=1}^{\kappa} m_n^2 >1$.

For small-amplitude waves,  we have $\sigma_n \approx \sqrt{g \; k_n} =\bar\sigma_n$, which leads to
\begin{equation}
  \left|\sum_{n=1}^{\kappa} m_n \; \bar{\sigma}_n\right| \approx \left|\sum_{n=1}^{\kappa} m_n \; {\sigma}_n\right| = \sigma' 
\end{equation}
Substituting the above expression and (\ref{def:k'}) into (\ref{resonance:general}),  we have the resonance condition (for $\kappa$ small-amplitude primary waves) in the form: 
\begin{equation}
g |{\bf k}'| = \sigma'^2, \nonumber
\end{equation} 
i.e.
\begin{equation}
\sigma' =\sqrt{g |{\bf k}'|}.    \label{resonance:general:small}
\end{equation}
The above resonance condition clearly reveals the physical relationship between the nonlinear-interaction wavenumber ${\bf k}' = \sum\limits_{n=1}^{\kappa} m_n \; {\bf k}_n$ and the nonlinear-interaction angular frequency $\sigma' = \left|   \sum\limits_{n=1}^{\kappa} m_n \; \sigma_n  \right|$.   

Mathematically,  the wave resonance condition (\ref{resonance:general:small}) can be derived in the frame of the HAM.  Note that  the HAM provides us great freedom to chose the auxiliary linear operator.    The auxiliary linear operator $\cal L$ defined by (\ref{def:L:general})  contains the term $\bar\sigma_i = \sqrt{g |{\bf k}_i|}$ which has physical meaning only for small-amplitude waves.  So, for large-amplitude waves, we should replace the term $\bar\sigma_i = \sqrt{g |{\bf k}_i|}$ in (\ref{def:L:general}) by the given angular frequency $\sigma_i$.  In other words,  for primary waves with large-amplitudes, we should choose the auxiliary linear operator
\begin{equation}
{\cal L}\phi = \left( \sum_{m=1}^{\kappa}\sum_{n=1}^{\kappa} \sigma_m \; \sigma_n \frac{\partial^2 \phi}{\partial \xi_m \partial \xi_n} \right)+ g \frac{\partial^2 \phi}{\partial z^2},  \label{def:L:general:large}
\end{equation}
where $\sigma_i$ is the angular frequency of the $i$th primary wave.   In fact, the above linear operator comes from the linear part of the nonlinear boundary condition (\ref{bc:varphi}).  The eigenvalue of the above linear operator reads
\begin{equation}
\lambda_{m_1,m_2,\cdots,m_\kappa} = g \left|\sum_{n=1}^{\kappa} m_n {\bf k}_n\right| - \left(  \sum_{n=1}^{\kappa} m_n \sigma_n \right)^2 \label{def:eigenvalue:general:large}.
\end{equation} 
Enforcing the above formula to be zero,  we obtain 
\begin{equation}  
\left|   \sum_{n=1}^{\kappa} m_n \sigma_n \right| = \sqrt{g \left|\sum_{n=1}^{\kappa} m_n {\bf k}_n\right|},      
\end{equation}
which is   exactly the wave resonance condition $\sigma' = \sqrt{g {\bf k}'}$  defined by  (\ref{resonance:general:small}).

Let $\sigma'_0$ denote the  ``natural'' angular frequency  of a single traveling wave with the  wavenumber ${\bf k}'$ and the  wave amplitude $a'$.    In case of small wave amplitudes,  according to the linear theory, we have the ``natural'' angular frequency  $\sigma'_0  \approx  \sqrt{g \; |{\bf k}'|}$.   Then, the above resonance condition becomes
\begin{equation}
\sigma' = \sigma'_0,  \label{resonance:general:large:A}
\end{equation}
i.e.
\begin{equation}
 \left|\sum_{n=1}^{\kappa} m_n \; {\sigma}_n\right| = \sigma'_0 , \;\;  \hspace{1.0cm} \sum_{n=1}^{\kappa} m_n^2 >1.  \label{resonance:general:large:B}
\end{equation}
Physically speaking, the wave resonance occurs when the nonlinear-interaction angular frequency $\sigma' = \left|\sum\limits_{n=1}^{\kappa} m_n \; {\sigma}_n\right|$ of the corresponding nonlinear-interaction wave with wavenumber ${\bf k}' = \sum\limits_{n=1}^{\kappa} m_n \; {\bf k}_n$ equals to its ``natural'' angular frequency $\sigma'_0$.     Note that,  different from the nonlinear-interaction angular frequency  $\sigma'$  that  is  a kind of sum of angular frequencies of primary waves,  the  ``natural'' angular frequency $\sigma'_0$ of the corresponding wave number  ${\bf k}'$ depends  only upon  the wavenumber ${\bf k}'$  and its amplitude $a'$, but has nothing to do with the angular frequencies of primary waves.     Thus,   in general,  the nonlinear-interaction angular frequency $\sigma'$ is not equal to  the ``natural'' angular frequency $\sigma'_0$ of the nonlinear-interaction wave with wavenumber ${\bf k}'$.    So, the wave resonance condition is indeed rather special.   This physical explanation agrees well with the traditional resonance theory.    So, (\ref{resonance:general:large:A}) and (\ref{resonance:general:large:B}) reveal the physical essence of  the gravity wave resonance.   

Although (\ref{resonance:general:large:B})  is derived from the resonance condition (\ref{resonance:general}) for {\em small} wave amplitudes, this physical mechanism of gravity wave resonance has general meanings and holds for large wave amplitudes even if $\sigma_n \approx \sqrt{g\; k_n}$ is not a good approximation.   So, (\ref{resonance:general:large:B})  is  also  the wave resonance condition for arbitrary number of primary  waves with {\em large} amplitudes.    It should be emphasized that the wave resonance condition (\ref{resonance:general:large:B})  logically contains the resonance condition (\ref{resonance:general}) for arbitrary number of small-amplitude waves and Phillips'  resonance condition (\ref{resonance:Phillips}) for four small-amplitude waves.   Thus, it is rather general.  
  
When the wave resonance condition  (\ref{resonance:general:large:B}) is satisfied and  the wave energy transfers from the primary waves to a resonant one,  the amplitudes of primary waves decreases and the amplitude of the resonant wave increases.   Therefore, the angular frequencies $\sigma_n$ of each primary waves decrease but the ``natural'' angular frequency of the resonant wave increases so that the resonance condition (\ref{resonance:general:large:B})  does not hold any more.    As a result,   the ``natural'' frequency $\sigma'_0$ departs more and more from the  nonlinear-interaction frequency  $\sigma'$,  and  the nonlinear-interaction wave  gains less and less energy from the primary waves, until the whole wave system is in equilibrium.    This explains why a resonant gravity wave has finite value of amplitude.    As mentioned above,   a resonant simple pendulum acted by an external force with the phase difference $\pi$, as shown in Fig.\ref{figure:pendulum},  loses its energy so that the maximum angel of oscillation $\theta_{max}$ decreases.  Similarly,   when the wave resonance condition  (\ref{resonance:general:large:B})  is satisfied, it is also possible that the wave energy transfers from the resonant wave to primary ones  so that the amplitude of resonant wave decreases and the amplitudes of primary waves  increase:  this explains why the amplitude of a resonant wave may be much smaller than those of primary ones, as shown in Table~\ref{Table:wave-energy:Group-1}.    
  
\section{Concluding remark and discussions}

The main findings and concluding remarks are outlined below.    

First of all, based on a analytic technique for strongly nonlinear problems, namely the homotopy analysis method (HAM), a multiple-variable technique is proposed and applied  to give convergent series solution of a fully developed system of arbitrary number of primary periodic traveling waves.   Different from perturbation techniques used by Phillips \cite{Phillips1960JFM} and Longuet-Higgins \cite{LonguetHiggins1962JFM}, this multiple-variable technique does not depend upon any small physical parameters,   and besides provides a convenient way to ensure the fast convergence of solution series.  By means of this multiple-variable technique, the time $t$ does not explicitly appear for a fully developed wave system:  this not only greatly simplifies solving the problem mathematically, but also contributes a lot to revealing  the  physical meanings clearly (some users of the HAM solved nonlinear wave-type PDEs by simply expanding the solution in Taylor series with respect to the time $t$.  Unfortunately, this often leads to very complicated solution expressions with rather little physical meanings).    Thus, the homotopy multiple-variable method has general meanings and can be widely applied to different types of nonlinear problems in science and engineering.    For example, although this method is used here for fully developed gravity waves, it can be applied to study the evolution of nonlinear waves far from equilibrium (such as the famous natural phenomena about freak wave \cite{Kharif2003, Paul2005, Paul2009}), as long as we introduce a ``slow''  time-scale $\tau = t/(1+t)$ as an additional variable to describe the time-dependent variation of the wave amplitude and angular frequency.       

Secondly, by means of the homotopy multiple-variable method,  we illustrated that the amplitudes of all wave components of a fully developed wave system are finite constants, even if the resonance condition is exactly satisfied.  Besides, we revealed, maybe for the first time,  that a fully developed resonant wave system may have multiple solutions.  Especially,  it is found that the amplitude of resonant waves might be much smaller than that of primary waves, and that a resonant wave may  contain only  the few of the whole wave energy.   These results differ from some of our traditional thoughts, but strongly suggest that, due to nonlinear interaction,  the evolution of a  multiple-wave system might be rather complicated.   At the end of \S 4, some physical explanations for these results are given.         

Third, by means of the homotopy multiple-variable method,  we derived  two general  wave resonance conditions (\ref{resonance:general})  and (\ref{resonance:general:large:B})   for arbitrary number of  primary periodic traveling  waves: the former holds for small-amplitude waves, but the latter works even for  large-amplitude waves.   These two  resonance conditions  logically contain Phillips' resonance condition (\ref{resonance:Phillips}) for four small-amplitude waves, and thus are more general.    Especially, the wave resonance condition (\ref{resonance:general:large:B})   opens a new way to study the strongly nonlinear  interactions of more than four primary traveling waves with large amplitudes.       

Mathematically, our computations suggest that, for a fully developed wave system with small amplitudes,  the main  wave energy distribute in the wave components whose eigenvalues to the linear operator (\ref{def:L:general}) or (\ref{def:L:general:large}) are zero (or close to zero).  
Physically speaking, the primary and resonant waves contain the main of the wave energy.   However,  as the wave amplitudes increases so that the nonlinearity becomes stronger,  the primary and resonant waves as a whole contain less and less percentage of wave energy, as shown in Fig.~\ref{figure:energy:Pi[0]}.  

There are some open questions.   The resonance condition (\ref{resonance:general:large:B})  for arbitrary number of traveling waves with large amplitudes is given from the physical view-points of resonance.    Although this resonance condition explains very well why the  amplitude of a resonant wave is finite and why it can be much smaller than those of primary ones, it should be verified by experiments or other analytical/numerical approaches.     Besides,  it is worthwhile studying the evolution of a system of multiple traveling waves far from equilibrium.   

Finally, it should be pointed out once again that the homotopy multiple-variable method proposed in this article is more general than the famous multiple-scales techniques in perturbation theory.  By means of the perturbation multiple-scale technique, one often rewrites a unknown function $f(t)$ in the form $F(T_0, T_1, T_2)$, where
\[   T_0 = t, \;\; T_1 =\epsilon \; t, \;\; T_2 = \epsilon^2 \; t \]
denote different timescales  with the small physical parameter $\epsilon$.  By means of this traditional multiple-scale technique, a nonlinear problem is often transformed into a sequence of linear perturbed problems via the small physical parameter $\epsilon$.    Using the homotopy multiple-variable method, we can also rewrite $f(t)$ by  $\check F(\xi_0, \xi_1,\xi_2) $  with the definition
\[   \xi_n = \epsilon^n \; t. \] 
However, different from the multiple-scale perturbation  techniques,  we now do {\em not} need any small physical parameters to transform the original nonlinear problem into a sequence of linear sub-problems.   Furthermore, it is easy to get high-order approximation by our approach, as illustrated in this article.   Especially, if the multiple-variables are properly defined with clear physical meanings,  this method is helpful to get results with important physical meanings.   This work illustrates that the homotopy multiple-variable method can overcome the restrictions of traditional analytic methods and besides it belongs to the times of computer.    It seems  that the homotopy multiple-variable technique can be applied widely to solve different types of  {\em strongly}  nonlinear problems in  science and engineering.

\vspace{0.5cm}

\setlength{\parindent}{0.0cm}{\bf Acknowledgements}    Thanks to Professor Roger Grimshaw (Loughborough University, UK) for some discussions about gravity wave resonance via emails, and  to Dr. Zhiliang Lin (Shanghai Jiaotong University) for his assistance in plotting the figure for the simple pendulum.  This work  is supported by National Natural Science Foundation of China  (Approve No. 10572095) and State Key Laboratory of Ocean Engineering (Approve No.  GKZD010002).

\vspace{0.50cm}

\newpage

\bibliography{liao,wave}

\bibliographystyle{unsrt}

\clearpage\newpage

\begin{center}
{\bf Appendix A}\\ {\bf The detailed derivation  of (\ref {bc:phi:m}) and (\ref {bc:eta:m})} 
\end{center}

\setlength{\parindent}{0.75cm}

Write 
\begin{equation}
\left(\sum_{i=1}^{+\infty}  \eta_i \; q^i \right)^m =\sum_{n=m}^{+\infty} \mu_{m,n} \; q^n, \label{def:mu:0}
\end{equation}
with the definition
\begin{equation}
\mu_{1,n}(\xi_1,\xi_2) = \eta_{n}(\xi_1,\xi_2), \;\; n\geq 1.   \label{def:mu:1}
\end{equation}
Then, 
\begin{eqnarray}
&&\left(\sum_{i=1}^{+\infty}  \eta_i \; q^i \right)^{m+1}  =  \left( \sum_{n=m}^{+\infty} \mu_{m,n} \; q^n \right) \left(\sum_{i=1}^{+\infty}  \eta_i \; q^i \right) \nonumber\\
&=& \sum_{s=m+1}^{+\infty} q^s \left( \sum_{n=m}^{s-1} \mu_{m,n} \; \eta_{s-n}\right) = \sum_{n=m+1}^{+\infty} q^n \left( \sum_{i=m}^{n-1} \mu_{m,i} \; \eta_{n-i}\right) \nonumber\\
&=& \sum_{j=m+1}^{+\infty} \mu_{m+1,n} \; q^n, 
\end{eqnarray}
which gives 
\[
\mu_{m+1,n}(\xi_1,\xi_2) = \sum_{i=m}^{n-1} \mu_{m,i}(\xi_1,\xi_2) \; \eta_{n-i}(\xi_1,\xi_2), \;\;\; m\geq 1, \; n\geq m+1,
\]
i.e.
\begin{equation}
\mu_{m,n}(\xi_1,\xi_2) = \sum_{i=m-1}^{n-1} \mu_{m-1,i}(\xi_1,\xi_2) \; \eta_{n-i}(\xi_1,\xi_2), \;\;\; m\geq 2, \; n\geq m. \label{def:mu:2}
\end{equation}
Thus, by means of (\ref{def:mu:1}) and (\ref{def:mu:2}),  one can easily get $\mu_{m,n}$ even for large $m$ and $n$.  

Define
\[      \psi_{i,j}^{n,m}(\xi_1,\xi_2)  =  \frac{\partial^{i+j}}{\partial \xi_1^i \partial \xi_2^j} \left(\frac{1}{m!} \; \left. \frac{\partial^m \phi_n}{\partial z^m} \right|_{z=0}\right).  \]
By Taylor series,  we have  for any $z$ that
\begin{equation}
    \phi_n(\xi_1,\xi_2,z) = \sum_{m=0}^{+\infty} \left(\left.\frac{1}{m!} \; \frac{\partial^m \phi_n}{\partial z^m}  \right|_{z=0}\right) z^m  = 
     \sum_{m=0}^{+\infty} \psi_{0,0}^{n,m} \; z^m \label{def:phi[n]:Taylor:0}
\end{equation}
and
\begin{equation}
    \frac{\partial^{i+j} \phi_n}{\partial \xi_1^i \partial \xi_2^j} = \sum_{m=0}^{+\infty} \frac{\partial^{i+j} }{\partial \xi_1^i \partial \xi_2^j} \left(\left.\frac{1}{m!} \; \frac{\partial^m \phi_n}{\partial z^m}  \right|_{z=0}\right) z^m  = 
     \sum_{m=0}^{+\infty} \psi_{i,j}^{n,m} \; z^m. \label{def:phi[n]:Taylor:1}
\end{equation}
Then, on $z = \check{\eta}(\xi_1,\xi_2;q) $,  we have using (\ref{def:mu:0}) that
\begin{eqnarray}
\frac{\partial^{i+j}\phi_n}{\partial \xi_1^i \partial\xi_2^j}
&=& \sum_{m=0}^{+\infty} \psi_{i,j}^{n,m}  \; \left( \sum_{s=1}^{+\infty} \eta_{s} \; q^s\right)^m \nonumber\\
&=& \psi_{i,j}^{n,0}+ \sum_{m=1}^{+\infty} \psi_{i,j}^{n,m} \; \left( \sum_{s=m}^{+\infty} \mu_{m,s} \; q^s \right)  \nonumber\\
&=& \psi_{i,j}^{n,0} +\sum_{s=1}^{+\infty}  q^s \left(  \sum_{m=1}^{s} \psi_{i,j}^{n,m} \; \mu_{m,s}\right)  \nonumber\\
&=& \psi_{i,j}^{n,0} +\sum_{m=1}^{+\infty}  q^m \left(  \sum_{s=1}^{m} \psi_{i,j}^{n,s} \; \mu_{s,m}\right)  \nonumber\\
&=&  \sum_{m=0}^{+\infty}\beta_{i,j}^{n,m}(\xi_1,\xi_2) \; q^m, \label{series:phi:n}
\end{eqnarray}
where
\begin{eqnarray}
\beta_{i,j}^{n,0} &=& \psi_{i,j}^{n,0},\label{def:beta:1}\\
\beta_{i,j}^{n,m} &=& \sum_{s=1}^{m} \psi_{i,j}^{n,s} \; \mu_{s,m}, \;\; m\geq 1, \label{def:beta:2}
\end{eqnarray}
i.e.
\begin{eqnarray}
\beta_{i,j}^{n,1} &=&  \eta_1\; \psi_{i,j}^{n,1}, \nonumber\\
\beta_ {i,j}^{n,2} &=&  \eta_2  \;\psi_ {i,j}^{n,1}  + \eta_1^2 \; \psi_ {i,j}^{n,2} , \nonumber\\
\beta_ {i,j}^{n,3} &=&  \eta_3 \; \psi_ {i,j}^{n,1}  + 2 \eta_1 \; \eta_2 \; \psi_ {i,j}^{n,2}  + \eta_1^3 \; \psi_ {i,j}^{n,3} , \nonumber\\
\beta_ {i,j}^{n,4} &=&   \eta_4 \; \psi_ {i,j}^{n,1}  +\left(2\eta_1\eta_3+\eta_2^2\right) \psi_ {i,j}^{n,2} + 3 \eta_1^2 \eta_2  \; \psi_ {i,j}^{n,3}+ \eta_1^4 \;\psi_ {i,j}^{n,4},\nonumber\\
&\vdots&  \nonumber
\end{eqnarray}   
Similarly,  on  $z = \check{\eta}(\xi_1,\xi_2;q) $, it holds
\begin{eqnarray}
\frac{\partial^{i+j}}{\partial\xi_1^i \partial\xi_2^j}\left(\frac{\partial \phi_n}{\partial z}\right) &=& \sum_{m=0}^{+\infty} \frac{\partial^{i+j}}{\partial\xi_1^i \partial\xi_2^j} \left(\left. \frac{1}{m!} \frac{\partial^{m+1}\phi_n}{\partial z^{m+1}} \right|_{z=0} \right)\; \check{\eta}^m\nonumber\\
&=& \sum_{m=0}^{+\infty} (m+1)\psi_ {i,j}^{n,m+1}\;\left( \sum_{s=1}^{+\infty} \eta_s \; q^s\right)^m\nonumber\\
&=& \sum_{m=0}^{+\infty} \gamma_ {i,j}^{n,m}(\xi_1,\xi_2)\; q^m,  \label{series:phi:n:z}
\end{eqnarray}
and
\begin{eqnarray}
\frac{\partial^{i+j}}{\partial\xi_1^i \partial\xi_2^j}\left(\frac{\partial^2 \phi_n}{\partial z^2}\right) &=& \sum_{m=0}^{+\infty} \frac{\partial^{i+j}}{\partial\xi_1^i \partial\xi_2^j} \left(\left. \frac{1}{m!} \frac{\partial^{m+2}\phi_n}{\partial z^{m+2}} \right|_{z=0}\right) \; \check{\eta}^m\nonumber\\
&=&\sum_{m=0}^{+\infty} (m+1)(m+2) \; \psi_ {i,j}^{n,m+2}\;\left( \sum_{s=1}^{+\infty} \eta_s \; q^s \right)^m\nonumber\\
&=& \sum_{m=0}^{+\infty} \delta_ {i,j}^{n,m}(\xi_1,\xi_2)\; q^m,  \label{series:phi:n:zz}
\end{eqnarray}
where
\begin{eqnarray}
\gamma_ {i,j}^{n,0}  &=& \psi_ {i,j}^{n,1}, \label{def:gamma:1}\\
\gamma_ {i,j}^{n,m} &=& \sum_{s=1}^{m} (s+1) \psi_ {i,j}^{n,s+1} \; \mu_{s,m}, \;\; m \geq 1,     \label{def:gamma:2}
\end{eqnarray}
i.e.
\begin{eqnarray}
\gamma_ {i,j}^{n,1} &=&  2\eta_1\; \psi_ {i,j}^{n,2}, \nonumber\\
\gamma_ {i,j}^{n,2} &=&  2\eta_2  \;\psi_ {i,j}^{n,2}  + 3\eta_1^2 \; \psi_ {i,j}^{n,3} , \nonumber\\
\gamma_ {i,j}^{n,3} &=&  2\eta_3 \; \psi_ {i,j}^{n,2}  + 6 \eta_1 \; \eta_2 \; \psi_ {i,j}^{n,3}  + 4\eta_1^3 \; \psi_ {i,j}^{n,4} , \nonumber\\
\gamma_ {i,j}^{n,4} &=&  2 \eta_4 \; \psi_ {i,j}^{n,2}  +3\left(2\eta_1\eta_3+\eta_2^2\right) \psi_ {i,j}^{n,3} + 12 \eta_1^2 \eta_2  \; \psi_ {i,j}^{n,4}+ 5\eta_1^4 \;\psi_ {i,j}^{n,5},\nonumber\\
&\vdots&  \nonumber
\end{eqnarray}
and
\begin{eqnarray}
\delta_ {i,j}^{n,0} &=& 2\psi_ {i,j}^{n,2},  \label{def:delta:1}\\
\delta_ {i,j}^{n,m} &=& \sum_{s=1}^{m} (s+1)(s+2) \psi_ {i,j}^{n,s+2} \; \mu_{s,m}, \;\; m \geq 1,      \label{def:delta:2}
\end{eqnarray}
i.e.
\begin{eqnarray}
\delta_ {i,j}^{n,1} &=&  6\eta_1\; \psi_ {i,j}^{n,3}, \nonumber\\
\delta_ {i,j}^{n,2} &=&  6\eta_2  \;\psi_ {i,j}^{n,3}  + 12\eta_1^2 \; \psi_ {i,j}^{n,4} , \nonumber\\
\delta_ {i,j}^{n,3} &=&  6\eta_3 \; \psi_ {i,j}^{n,3}  + 24 \eta_1 \; \eta_2 \; \psi_ {i,j}^{n,4}  + 20\eta_1^3 \; \psi_ {i,j}^{n,5} , \nonumber\\
\delta_ {i,j}^{n,4} &=&  6 \eta_4 \; \psi_ {i,j}^{n,3}  +12\left(2\eta_1\eta_3+\eta_2^2\right) \psi_ {i,j}^{n,4} + 60 \eta_1^2 \eta_2  \; \psi_ {i,j}^{n,5}+ 30\eta_1^4 \;\psi_ {i,j}^{n,6},\nonumber\\
&\vdots&  \nonumber
\end{eqnarray}
Note that the {\em  explicit}  expressions of $\beta_{i,j}^{n,m}(\xi_1,\xi_2), \gamma_{i,j}^{n,m}(\xi_1,\xi_2), \delta_{i,j}^{n,m}(\xi_1,\xi_2)$ defined above  can be easily obtained by symbolic software such as Mathematica, Maple and so on.    

Then, on $z = \check{\eta}(\xi_1,\xi_2;q) $, it holds  using (\ref{series:phi:n}) that
\begin{eqnarray}
\check{\phi}(\xi_1,\xi_2,\check{\eta};q) &=& \sum_{n=0}^{+\infty} \phi_n(\xi_1,\xi_2,\check{\eta}) \; q^n  
=\sum_{n=0}^{+\infty} q^n \left[\sum_{m=0}^{+\infty} \beta_{0,0}^{n,m}(\xi_1,\xi_2) \; q^m \right]\nonumber\\
&=&\sum_{n=0}^{+\infty}\sum_{m=0}^{+\infty} \beta_{0,0}^{n,m}(\xi_1,\xi_2)\; q^{m+n} =\sum_{s=0}^{+\infty}q^s \left[ \sum_{m=0}^{s} \beta_{0,0}^{s-m,m}(\xi_1,\xi_2) \right] \nonumber\\
&=&\sum_{n=0}^{+\infty} \bar{\phi}^{0,0}_n (\xi_1,\xi_2) \; q^n,\label{def:phi:surface}
\end{eqnarray}
where
\begin{eqnarray}
 \bar{\phi}^{0,0}_n (\xi_1,\xi_2)  = \sum_{m=0}^{n} \beta_{0,0}^{n-m,m}.  \label{def:phi:bar}
 \end{eqnarray}
Similarly, we have
\begin{eqnarray}
\frac{\partial ^{i+j}\check\phi}{\partial\xi_1^i \partial \xi_2^j} =  \sum_{n=0}^{+\infty} \bar{\phi}^{i,j}_n (\xi_1,\xi_2) \; q^n, \label{def:phi:bar:2}
\end{eqnarray}
where
\begin{eqnarray}
 \bar{\phi}^{i,j}_n (\xi_1,\xi_2)  = \sum_{m=0}^{n} \beta_{i,j}^{n-m,m}.  \label{def:phi:bar:3}
 \end{eqnarray}
Similarly, on $z = \check{\eta}(\xi_1,\xi_2;q) $,  we have
\begin{eqnarray}
\frac{\partial ^{i+j}}{\partial\xi_1^i \partial \xi_2^j} \left(\frac{\partial \check{\phi}}{\partial z}\right) &=& \sum_{n=0}^{+\infty} 
\bar{\phi}^{i,j}_{z,n}(\xi_1,\xi_2) \; q^n,\label{def:phi:z}\\
\frac{\partial ^{i+j}}{\partial\xi_1^i \partial \xi_2^j} \left(\frac{\partial^2 \check{\phi}}{\partial z^2} \right)&=& \sum_{n=0}^{+\infty} \bar{\phi}^{i,j}_{zz,n}(\xi_1,\xi_2) \; q^n, \label{def:phi:zz}
\end{eqnarray}  
where
\begin{eqnarray}
 \bar{\phi}^{i,j}_{z,n} (\xi_1,\xi_2)  &=& \sum_{m=0}^{n} \gamma_{i,j}^{n-m,m} , \label{def:phi:z:bar}\\
 \bar{\phi}^{i,j}_{zz,n} (\xi_1,\xi_2)  &=& \sum_{m=0}^{n} \delta_{i,j}^{n-m,m}. \label{def:phi:zz:bar}
 \end{eqnarray}

 Then, on $z =\check\eta(\xi_1,\xi_2;q)$,   it holds using (\ref{def:phi:bar:2}) and (\ref{def:phi:z}) that
 \begin{eqnarray}
 \check{f} &=& \frac{1}{2} \hat{\nabla} \check{\phi} \cdot   \hat{\nabla} \check{\phi}\nonumber\\
 &=&\frac{ k_1^2}{2}\left(\frac{\partial \check\phi}{\partial \xi_1}\right)^2 + {\bf k}_1 \cdot {\bf k}_2 \; \frac{\partial \check\phi}{\partial \xi_1} \; \frac{\partial \check\phi}{\partial \xi_2}  + \frac{k_2^2}{2} \left(\frac{\partial \check\phi}{\partial \xi_2}\right)^2+  \frac{1}{2}\left(\frac{\partial \check\phi}{\partial z}\right)^2\nonumber\\
 &=& \frac{ k_1^2}{2}\left(\sum_{m=0}^{+\infty}\bar{\phi}^{1,0}_m \; q^m\right)^2 + {\bf k}_1 \cdot {\bf k}_2 \; \left(\sum_{m=0}^{+\infty} \bar{\phi}^{1,0}_m  \; q^m\right) \left(\sum_{n=0}^{+\infty} \bar{\phi}^{0,1}_n \; q^n\right)
 \nonumber\\
 &+& \frac{k_2^2}{2} \left(\sum_{m=0}^{+\infty} \bar{\phi}^{0,1}_m \; q^m\right)^2+  \frac{1}{2}\left(\sum_{m=0}^{+\infty} \bar{\phi}^{0,0}_{z,m} \; q^m\right)^2\nonumber\\
 &=& \sum_{m=0}^{+\infty} \Gamma_{m,0}(\xi_1,\xi_2) \; q^m,  \label{def:f:bar}
 \end{eqnarray}
 where
 \begin{eqnarray}
 \Gamma_{m,0}(\xi_1,\xi_2) &=& \frac{k_1^2}{2} \sum_{n=0}^{m} \bar\phi_n^{1,0} \; \bar\phi_{m-n}^{1,0}+  {\bf k}_1 \cdot {\bf k}_2 \sum_{n=0}^{m} \bar\phi_n^{1,0} \bar\phi_{m-n}^{0,1}\nonumber\\
 &&+ \frac{k_2^2}{2} \sum_{n=0}^{m} \bar\phi_n^{0,1} \bar\phi_{m-n}^{0,1}+\frac{1}{2} \sum_{n=0}^{m} \bar\phi_{z,n}^{0,0} \; \bar\phi_{z,m-n}^{0,0}. \label{def:Gamma}
 \end{eqnarray}
 
Similarly, on $z =\check\eta(\xi_1, \xi_2;q)$,  it holds
\begin{eqnarray}
\frac{\partial \check f}{\partial \xi_1} &=& \hat\nabla \check\phi \cdot   \hat\nabla \left(\frac{\partial \check\phi}{\partial \xi_1}\right)\nonumber\\
&=& k_1^2 \frac{\partial \check\phi}{\partial \xi_1} \frac{\partial^2 \check\phi}{\partial \xi_1^2} +k_2^2 \frac{\partial \check\phi}{\partial \xi_2} \frac{\partial^2 \check\phi}{\partial \xi_1 \partial \xi_2} + \frac{\partial \check\phi}{\partial z} \frac{\partial }{\partial \xi_1}\left( \frac{\partial \check\phi}{\partial z}\right) \nonumber\\
&+& {\bf k}_1 \cdot {\bf k}_2  \left(  \frac{\partial \check\phi}{\partial \xi_1} \frac{\partial^2 \check\phi}{\partial \xi_1 \partial \xi_2} +  \frac{\partial \check\phi}{\partial \xi_2} \frac{\partial^2 \check\phi}{\partial \xi_1^2}\right)\nonumber\\
&=& k_1^2 \left( \sum_{m=0}^{+\infty} \bar\phi_m^{1,0} \; q^m\right) \left( \sum_{n=0}^{+\infty} \bar\phi_n^{2,0} \; q^n\right)+k_2^2 \left( \sum_{m=0}^{+\infty} \bar\phi_m^{0,1} \; q^m\right) \left( \sum_{n=0}^{+\infty} \bar\phi_n^{1,1} \; q^n\right) \nonumber\\
&+& \left( \sum_{m=0}^{+\infty} \bar\phi_{z,m}^{0,0} \; q^m\right) \left( \sum_{n=0}^{+\infty} \bar\phi_{z,n}^{1,0} \; q^n\right)
+ {\bf k}_1 \cdot {\bf k}_2  \left( \sum_{m=0}^{+\infty} \bar\phi_m^{1,0} \; q^m\right) \left( \sum_{n=0}^{+\infty} \bar\phi_n^{1,1} \; q^n\right)\nonumber\\
&+&  {\bf k}_1 \cdot {\bf k}_2  \left( \sum_{m=0}^{+\infty} \bar\phi_m^{2,0} \; q^m\right) \left( \sum_{n=0}^{+\infty} \bar\phi_n^{0,1} \; q^n\right)\nonumber\\
&=&\sum_{m=0}^{+\infty} \Gamma_{m,1}(\xi_1,\xi_2) \; q^m, \label{def:Gamma:x1}
\end{eqnarray}
where
\begin{eqnarray}
\Gamma_{m,1}(\xi_1,\xi_2) &=&  \sum_{n=0}^{m} \left( k_1^2 \; \bar\phi_n^{1,0}  \; \bar\phi_{m-n}^{2,0} 
+k_2^2   \; \bar\phi_n^{0,1} \; \bar\phi_{m-n}^{1,1} +  \bar\phi_{z,n}^{0,0} \; \bar\phi_{z,m-n}^{1,0}\right) \nonumber\\
&+& {\bf k}_1 \cdot {\bf k}_2 \sum_{n=0}^{m} \left(  \bar\phi_n^{1,0} \; \bar\phi_{m-n}^{1,1} +  \bar\phi_n^{2,0} \; \bar\phi_{m-n}^{0,1}\right),
\label{def:GAMMA:x1:coefficient}
\end{eqnarray}
and
\begin{eqnarray}
\frac{\partial \check f}{\partial \xi_2}  &=& \hat\nabla \check\phi \cdot   \hat\nabla \left(\frac{\partial \check\phi}{\partial \xi_2}\right)\nonumber\\
&=& k_1^2 \frac{\partial \check\phi}{\partial \xi_1} \frac{\partial^2 \check\phi}{\partial \xi_1 \partial \xi_2} +k_2^2 \frac{\partial \check\phi}{\partial \xi_2} \frac{\partial^2 \check\phi}{ \partial \xi_2^2} + \frac{\partial \check\phi}{\partial z} \frac{\partial }{\partial \xi_2}\left( \frac{\partial \check\phi}{\partial z}\right) \nonumber\\
&+& {\bf k}_1 \cdot {\bf k}_2  \left(  \frac{\partial \check\phi}{\partial \xi_1} \frac{\partial^2 \check\phi}{ \partial \xi_2^2} +  \frac{\partial \check\phi}{\partial \xi_2} \frac{\partial^2 \check\phi}{\partial \xi_1 \partial \xi_2}\right)\nonumber\\
&=& \sum_{m=0}^{+\infty} \Gamma_{m,2}(\xi_1,\xi_2) \; q^m,  \label{def:Gamma:x2}
\end{eqnarray}
where
\begin{eqnarray}
\Gamma_{m,2}(\xi_1,\xi_2) &=&  \sum_{n=0}^{m} \left( k_1^2 \; \bar\phi_n^{1,0} \; \bar\phi_{m-n}^{1,1} 
+k_2^2  \; \bar\phi_n^{0,1} \; \bar\phi_{m-n}^{0,2} +  \bar\phi_{z,n}^{0,0} \; \bar\phi_{z,m-n}^{0,1}\right) \nonumber\\
&+& {\bf k}_1 \cdot {\bf k}_2 \sum_{n=0}^{m} \left(  \bar\phi_n^{1,0}  \; \bar\phi_{m-n}^{0,2} +  \bar\phi_n^{0,1}  \; \bar\phi_{m-n}^{1,1}\right).
\label{def:GAMMA:x2:coefficient}
\end{eqnarray}
Besides,  on $z  =\check\eta(\xi_1,\xi_2;q)$, we have by means of (\ref{def:phi:bar:2}), (\ref{def:phi:z}) and (\ref{def:phi:zz}) that 
 \begin{eqnarray}
 \frac{\partial \check f}{\partial z} &=&  \hat\nabla \check \phi \cdot \hat\nabla \left(\frac{\partial \check{\phi}}{\partial z}\right) \nonumber\\
 &=& k_1^2 \; \frac{\partial \check\phi}{\partial \xi_1} \frac{\partial}{\partial \xi_1} \left( \frac{\partial \check\phi}{\partial z}\right) + k_2^2 \; \frac{\partial \check\phi}{\partial \xi_2} \frac{\partial}{\partial \xi_2} \left( \frac{\partial \check\phi}{\partial z}\right)+\frac{\partial \check\phi}{\partial z}  \frac{\partial^2 \check\phi}{\partial z^2}  \nonumber\\
 &+& {\bf k}_1 \cdot {\bf k}_2 \; \left[\frac{\partial \check\phi}{\partial \xi_1} \frac{\partial}{\partial \xi_2} \left( \frac{\partial \check\phi}{\partial z}\right) +\frac{\partial \check\phi}{\partial \xi_2} \frac{\partial}{\partial \xi_1} \left( \frac{\partial \check\phi}{\partial z}\right)  \right] \nonumber\\
 &=& \sum_{m=0}^{+\infty} \Gamma_{m,3}(\xi_1,\xi_2) \; q^m, \label{def:Gamma:z}
 \end{eqnarray}
 where
\begin{eqnarray}
\Gamma_{m,3}(\xi_1,\xi_2) &=&  \sum_{n=0}^{m} \left( k_1^2 \; \bar\phi_n^{1,0} \; \bar\phi_{z,m-n}^{1,0} 
+k_2^2  \; \bar\phi_n^{0,1} \; \bar\phi_{z,m-n}^{0,1} +  \bar\phi_{z,n}^{0,0} \; \bar\phi_{zz,m-n}^{0,0}\right) \nonumber\\
&+& {\bf k}_1 \cdot {\bf k}_2 \sum_{n=0}^{m} \left(  \bar\phi_n^{1,0}  \; \bar\phi_{z,m-n}^{0,1} +  \bar\phi_n^{0,1}  \; \bar\phi_{z,m-n}^{1,0}\right).
\label{def:GAMMA:z:coefficient}
\end{eqnarray}
Furthermore, using (\ref{def:phi:bar:2}), (\ref{def:Gamma:x1}), (\ref{def:Gamma:x2}) and (\ref{def:Gamma:z}),  we have
\begin{eqnarray}
\hat\nabla \check\phi \cdot \hat\nabla \check f &=& k_1^2 \frac{\partial \check\phi}{\partial\xi_1} \frac{\partial \check f}{\partial \xi_1} + k_2^2 \frac{\partial \check\phi}{\partial\xi_2} \frac{\partial \check f}{\partial \xi_2} + \frac{\partial \check\phi}{\partial z} \frac{\partial \check f}{\partial z}\nonumber\\
&+& {\bf k}_1 \cdot {\bf k}_2 \left( \frac{\partial \check\phi}{\partial\xi_1} \frac{\partial \check f}{\partial \xi_2} + \frac{\partial \check\phi}{\partial\xi_2} \frac{\partial \check f}{\partial \xi_1}\right)\nonumber\\
&=& \sum_{m=0}^{+\infty} \Lambda_m(\xi_1,\xi_2) \; q^m,  \label{def:Lambda}
\end{eqnarray}
where
\begin{eqnarray}
\Lambda_{m}(\xi_1,\xi_2) &=&  \sum_{n=0}^{m} \left( k_1^2 \; \bar\phi_n^{1,0} \; \Gamma_{m-n,1} 
+k_2^2  \; \bar\phi_n^{0,1} \; \Gamma_{m-n,2} +  \bar\phi_{z,n}^{0,0} \; \Gamma_{m-n,3}\right) \nonumber\\
&+& {\bf k}_1 \cdot {\bf k}_2 \sum_{n=0}^{m} \left(  \bar\phi_n^{1,0}  \; \Gamma_{m-n,2}+  \bar\phi_n^{0,1}  \; \Gamma_{m-n,1}\right).
\label{def:Lambda:coefficient}
\end{eqnarray}

 Then, using  (\ref {def:phi:bar:2}), (\ref{def:phi:z}), (\ref{def:Gamma:x1}), (\ref{def:Gamma:x2}) and (\ref{def:Lambda}),  we have on $z =\check\eta(\xi_1,\xi_2;q)$ that
   \begin{eqnarray}
&&{\cal N}\left[  \check\phi(\xi_1,\xi_2,z;q)\right] \nonumber\\
&=& \sigma_1^2 \; \frac{\partial^2 \check\phi}{\partial \xi_1^2}+2 \sigma_1 \sigma_2 \; \frac{\partial^2 \check\phi}{\partial \xi_1 \partial \xi_2}+ \sigma_2^2 \; \frac{\partial^2 \check\phi}{\partial \xi_2^2}+g \frac{\partial \check\phi}{\partial z} \nonumber\\
 && -2\left( \sigma_1 \;  \frac{\partial \check f}{\partial \xi_1} + \sigma_2 \; \frac{\partial \check f}{\partial \xi_2}\right) + \hat{\nabla} \check\phi \cdot  \hat{\nabla}  \check f \nonumber\\
 &=&  \sum_{m=0}^{+\infty} \Delta^\phi_m(\xi_1,\xi_2) \; q^m, \label{series:N:q}
  \end{eqnarray}
  where
  \begin{eqnarray}
  \Delta^\phi_m(\xi_1,\xi_2) &=& \sigma_1^2 \;\bar\phi_{m}^{2,0}+2 \sigma_1\sigma_2 \; \bar\phi_m^{1,1} + \sigma_{2}^2 \; \bar\phi_m^{0,2} +g \bar\phi_{z,m}^{0,0}\nonumber\\
  &-&2 \left(\sigma_{1} \;\Gamma_{m,1}+  \sigma_{2} \;  \Gamma_{m,2} \right) +\Lambda_m  \label{def:Delta:phi}
  \end{eqnarray}
for $ m\geq 0 $.

Using (\ref{series:phi:q}) and (\ref{series:phi:n}), we have  on $z=\check\eta(\xi_1,\xi_2;q)$ that
\begin{eqnarray}
\check\phi-\phi_0 &=& \sum_{n=1}^{+\infty} \phi_n(\xi_1,\xi_2,\check\eta) \; q^n  = \sum_{n=1}^{+\infty} q^n 
\left(\sum_{m=0}^{+\infty} \beta_{0,0}^{n,m} \; q^m \right)\nonumber\\
&=&\sum_{n=1}^{+\infty} q^n \left( \sum_{m=0}^{n-1}\beta_{0,0}^{n-m,m}\right)
\end{eqnarray}
and similarly
\begin{eqnarray}
\frac{\partial}{\partial z}\left(\check\phi-\phi_0\right) &=& \sum_{n=1}^{+\infty} \frac{\partial \phi_n}{\partial z} \; q^n  = \sum_{n=1}^{+\infty} q^n \left(\sum_{m=0}^{+\infty} \gamma_{0,0}^{n,m} \; q^m \right)\nonumber\\
&=&\sum_{n=1}^{+\infty} q^n \left( \sum_{m=0}^{n-1}\gamma_{0,0}^{n-m,m}\right),
\end{eqnarray}
respectively.    Then, on $z=\check\eta(\xi_1,\xi_2;q)$, it holds due to the linear property of the operator (\ref{def:L}) that
\begin{equation}
{\cal L}\left( \check \phi -\phi_0 \right) = \sum_{n=1}^{+\infty} S_n(\xi_1,\xi_2)\; q^n,
\end{equation}
where
\begin{equation}
S_n(\xi_1,\xi_2) = \sum_{m=0}^{n-1}\left( \bar\sigma_1^2 \; \beta_{2,0}^{n-m,m}
+2\bar\sigma_1 \bar\sigma_2  \; \beta_{1,1}^{n-m,m}
+ \bar\sigma_2^2\;  \beta_{0,2}^{n-m,m}+ g \; \gamma_{0,0}^{n-m,m}\right) . \label{def:S}
\end{equation}
Then, on $z=\check\eta(\xi_1,\xi_2;q)$, it holds
\begin{eqnarray}
(1-q) {\cal L}\left( \check \phi -\phi_0 \right) = (1-q)\sum_{n=1}^{+\infty} S_n\; q^n =\sum_{n=1}^{+\infty} \left( S_n-\chi_n \; S_{n-1}\right) q^n,\label{geq:zero:left}
\end{eqnarray}
where 
\begin{equation}
\chi_n = \left\{  
\begin{array}{cc}
0, & \mbox{when $n\leq 1$}, \\
1, & \mbox{when $n > 1$} . 
\end{array} \right.
\end{equation}
Substituting (\ref{geq:zero:left}), (\ref{series:N:q}) into (\ref{geq:zero}) and equating the like-power of $q$, we have the boundary condition: 
\begin{equation}
S_m(\xi_1,\xi_2)-\chi_m \; S_{m-1}(\xi_1,\xi_2) = c_0 \; \Delta^\phi_{m-1}(\xi_1,\xi_2),\;\;\;\;   m\geq 1. \label{bc:phi:mth:0}
\end{equation}
Define
\begin{equation}
\bar{S}_n(\xi_1,\xi_2) = \sum_{m=1}^{n-1}\left( \bar\sigma_1^2 \; \beta_{2,0}^{n-m,m} +2\bar\sigma_1 \bar\sigma_2 \;\beta_{1,1}^{n-m,m}
+ \bar\sigma_2^2  \;  \beta_{0,2}^{n-m,m}+  g \; \gamma_{0,0}^{n-m,m}\right) . \label{def:S:bar}
\end{equation}
Then,
\begin{eqnarray}
S_n &=& \left( \bar\sigma_1^2 \; \beta_{2,0}^{n,0} 
+2\bar\sigma_1\bar\sigma_2 \; \beta_{1,1}^{n,0}
+  \bar\sigma_2^2 \;  \beta_{0,2}^{n,0} +  g\;\gamma_{0,0}^{n,0}  \right) +\bar{S}_n\nonumber\\
&=& \left.  \left( \bar\sigma_1^2 \;\frac{\partial^2 \phi_n}{\partial \xi_1^2}
+2\bar\sigma_1 \bar\sigma_2  \; \frac{\partial^2 \phi_n}{\partial \xi_1 \partial \xi_2}
+ \bar\sigma_2^2 \; \frac{\partial^2 \phi_n}{\partial \xi_2^2}+ g \frac{\partial \phi_n}{\partial z}\right)\right|_{z=0} + \bar{S}_n
\end{eqnarray}
Substituting the above expression into (\ref{bc:phi:mth:0}) gives the boundary condition  on $z=0$: 
\begin{equation}
\bar{\cal L}\left( \phi_m \right)  =  c_0 \; \Delta^\phi_{m-1} +\chi_m\; S_{m-1} -\bar{S}_m,  \;\;\;  m \geq 1, \label{bc:phi:m:0}
\end{equation}
where $\bar{\cal L}$ is defined by (\ref{def:L:z=0}). 

 Substituting the series (\ref{series:eta:q}), (\ref{def:phi:bar:2}) and (\ref{def:f:bar}) into (\ref{bc:eta:zero}), equating the like-power of $q$, we have  
 \begin{equation}
 \eta_m(\xi_1,\xi_2) =   c_0  \; \Delta_{m-1}^\eta + \chi_m \; \eta_{m-1}, \;\;\;  m\geq 1, 
 \end{equation}
where
\[   \Delta_{m}^\eta = \eta_{m} - \frac{1}{g} \left( \sigma_{1} \; \bar\phi_{m}^{1,0}
 + \sigma_{2}  \; \bar\phi_{m}^{0,1}  - \Gamma_{m,0}\right).    \]

\newpage
\begin{center}
{\bf Appendix B} \\

{\bf A brief proof of  the Convergence Theorem}
\end{center}

\hspace{-0.7cm}Proof:  The potential function $\phi$ is a linear combination of the eigenfunctions  \[\Psi_{m,n} = \exp(|m{\bf k}_1+n{\bf k}_2|z) \; \sin(m\xi_1+n\xi_2)\]
for integers $m$ and $n$.  Thus, $\phi$ automatically satisfies the linear Laplace equation (\ref{geq:phi}) and the boundary condition (\ref{bc:phi:bottom}).  

If $\sum\limits_{m=0}^{+\infty} \Delta_m^\eta =0$, then
\begin{eqnarray}
\sum_{m=1}^{+\infty} \eta_m =\frac{1}{g}\left[ \sigma_1 \left(\sum_{m=0}^{+\infty}\bar{\phi}_m^{1,0}\right)+ \sigma_2 \left(\sum_{m=0}^{+\infty}\bar{\phi}_m^{0,1}\right)-\left(\sum_{m=0}^{+\infty}\Gamma_{m,0}\right)\right]. \nonumber
\end{eqnarray}
Setting $q=1$ in (\ref{def:phi:surface}) and (\ref{def:phi:bar:2}) gives on $z=\check\eta(\xi_1,\xi_2;1)$ that 
\[  \sum_{m=0}^{+\infty}\bar{\phi}_m^{1,0} =\frac{\partial \check\phi(\xi_1,\xi_2,z;1)}{\partial \xi_1}  , \;\;\;  \sum_{m=0}^{+\infty}\bar{\phi}_m^{0,1} =\frac{\partial \check\phi(\xi_1,\xi_2,z;1)}{\partial \xi_2}.   \]
Similarly, setting $q=1$ in (\ref{def:f:bar}) gives on $z=\check\eta(\xi_1,\xi_2;1)$ that 
\[  \sum_{m=0}^{+\infty}\Gamma_{m,0}  =  \frac{1}{2} \hat{\nabla} \check{\phi}(\xi_1,\xi_2,z;1) \cdot   \hat{\nabla} \check{\phi}(\xi_1,\xi_2,z;1).  \]
Thus, we have on $z=\check\eta(\xi_1,\xi_2;1)$ that 
\begin{eqnarray}
  &&\check\eta(\xi_1,\xi_2;1) \nonumber\\
  & = &\frac{1}{g}\left[\sigma_1 \frac{\partial \check\phi(\xi_1,\xi_2,z;1)}{\partial \xi_1} +\sigma_2 \frac{\partial \check\phi(\xi_1,\xi_2,z;1)}{\partial \xi_2}-  \frac{1}{2} \hat{\nabla} \check{\phi}(\xi_1,\xi_2,z;1) \cdot   \hat{\nabla} \check{\phi}(\xi_1,\xi_2,z;1) \right] . \hspace{1.2cm}  \label{eq:eta}
  \end{eqnarray}

Besides, when $\sum\limits_{m=0}^{+\infty} \Delta_m^\phi =0$,  setting $q=1$ in (\ref{series:N:q}) gives  
\begin{equation}
\sum_{m=0}^{+\infty}\Delta_m^\phi ={\cal N}\left[ \check\phi(\xi_1,\xi_2,z;1)\right]   = 0 \label{eq:phi}
\end{equation}
on the free surface $z=\check\eta(\xi_1,\xi_2;1)$.    Note that, according to (\ref{series:phi:q}) and (\ref{series:eta:q}),   $\check\phi(\xi_1,\xi_2,z;1)$ and $\check\eta(\xi_1,\xi_2;1)$ denote the homotopy-series  (\ref{series:phi}) and (\ref{series:eta}),  respectively.   Note also that (\ref{eq:eta}) and (\ref{eq:phi}) are exactly the two boundary conditions on the free surface.  Therefore, the homotopy-series  (\ref{series:phi}) and (\ref{series:eta}) satisfy the original governing equation (\ref{geq:phi}) and all boundary conditions (\ref{bc:eta}),(\ref{bc:phi}) and  (\ref{bc:phi:bottom}).  This ends the proof.  

\end{document}